\documentclass[a4paper,11pt]{article}
\usepackage{jheppub}
\usepackage[T1]{fontenc} 
\usepackage{amsmath}
\usepackage[caption=false]{subfig}

\usepackage{breqn}

\def\eps{\epsilon}
\def\cN{  {\cal N}  }
\def\cO{  {\cal O}  }
\newcommand{\bea}{\begin{eqnarray}}
\newcommand{\eea}{\end{eqnarray}}
\newcommand{\dr}{{\rm d}}
\newcommand{\nn}{\nonumber}

\allowdisplaybreaks[4]

\title{Analytic results for two-loop master integrals for Bhabha scattering  I}

\author[a]{Johannes M.\ Henn}
\author[b]{Vladimir A.\ Smirnov}
\affiliation[a]{Institute for Advanced Study, Princeton, NJ 08540, USA}
\affiliation[b]{Skobeltsyn Institute of Nuclear Physics of Moscow State University,
Moscow 119992, Russia}
\emailAdd {jmhenn@ias.edu}
\emailAdd {smirnov@theory.sinp.msu.ru}

\abstract{
We evaluate analytically the master integrals for one of two types of planar families 
contributing to massive two-loop Bhabha scattering in QED.
As in our previous paper, we apply a recently suggested new strategy to solve differential equations
for master integrals for families of Feynman integrals.
The crucial point of this strategy is to use a new basis of the master integrals
where all master integrals 
are pure functions of uniform weight.
This allows to cast the differential equations into a simple canonical form, 
which can straightforwardly be integrated order by order in $\eps$.
The boundary conditions are also particularly transparent in this setup.
We identify the class of functions relevant to this problem to all orders in $\eps$.
We present the results up to weight four for all except one integrals in terms of a 
subset of Goncharov polylogarithms, which one may call two-dimensional harmonic polylogarithms.
For one integral, and more generally at higher weight, the solution is written in terms of Chen iterated integrals.
}


\begin{document}

\maketitle
\flushbottom

\section{Introduction}  

Two-loop Feynman integrals contributing to Bhabha scattering were extensively studied because
of the corresponding high experimental precision achieved at $e^+ e^{-}$ colliders.
Planar two-loop box integrals with one-loop insertion were calculated in \cite{Bonciani:2003cj} 
by the method of differential equations (DE) \cite*{Kotikov:1990kg,Remiddi:1997ny,Gehrmann:1999as}, 
see \cite{Argeri:2007up,Smirnov:2012gma} for reviews.
Then all the two-loop integrals involving a closed fermion loop were evaluated in~\cite{Bonciani:2004gi}.
The first result for a massive two-loop double box was obtained 
in~\cite{Smirnov:2001cm} by the method of Mellin--Barnes (MB) representation, 
see chapter~5 of \cite{Smirnov:2012gma} for a review.
Then a similar result for a master integral with a numerator was obtained
\cite{Heinrich:2004iq} and pole parts in the dimensional regularization parameter 
$\epsilon=(4-D)/2$ for some other master integrals were evaluated.

An attempt to evaluate all the master integrals for Bhabha scattering, i.e. integrals belonging
to two families of planar diagrams and one family of non-planar diagrams was undertaken in
\cite*{Czakon:2004wm,Czakon:2005jd,Czakon:2005gi,Czakon:2006pa,Czakon:2006hb}, with the help of DE.
\footnote{Analytic results for other mass configurations were also obtained using this method, see e.g. \cite*{Anastasiou:2006hc,Czakon:2008ii,vonManteuffel:2013uoa} and references therein.}
However, starting from master integrals with five propagators,
the resulting DE involved highly non-trivial coupled sets of equations,
which complicated the situation considerably.
In those cases it was also found that the determination of the boundary constants 
was a complicated task. Finally, it remained unclear what precise class of functions
would be sufficient to express all master integrals.

One should mention that for phenomenological applications to Bhabha scattering
sometimes the small mass limit is sufficient, and this was investigated in
\cite*{Actis:2006dj,Actis:2007gi,Actis:2007pn,Actis:2007fs,Actis:2007zz,Actis:2008br}.
However, the 
problem of analytical evaluation of all the master integrals for Bhabha scattering
remained open up to now. 
Moreover, its analytic expressions are obviously desirable
in order to have control over errors in approximations, 
especially at intermediate energies  where the latter may not always be justified.
Furthermore, understanding the structure of loop integrals more generally 
is an interesting and important challenge. For example, until now it was not 
known which class of special functions one ultimately needs in order to describe
the $\eps$ expansion of the integrals. This is an important question, and, as we will see, 
understanding the answer to it will turn out to be almost tantamount to obtaining the full 
analytic solution.

\begin{figure}[t] 
\captionsetup[subfigure]{labelformat=empty}
\begin{center}
\subfloat[(1)]{\includegraphics[width=0.3\textwidth]{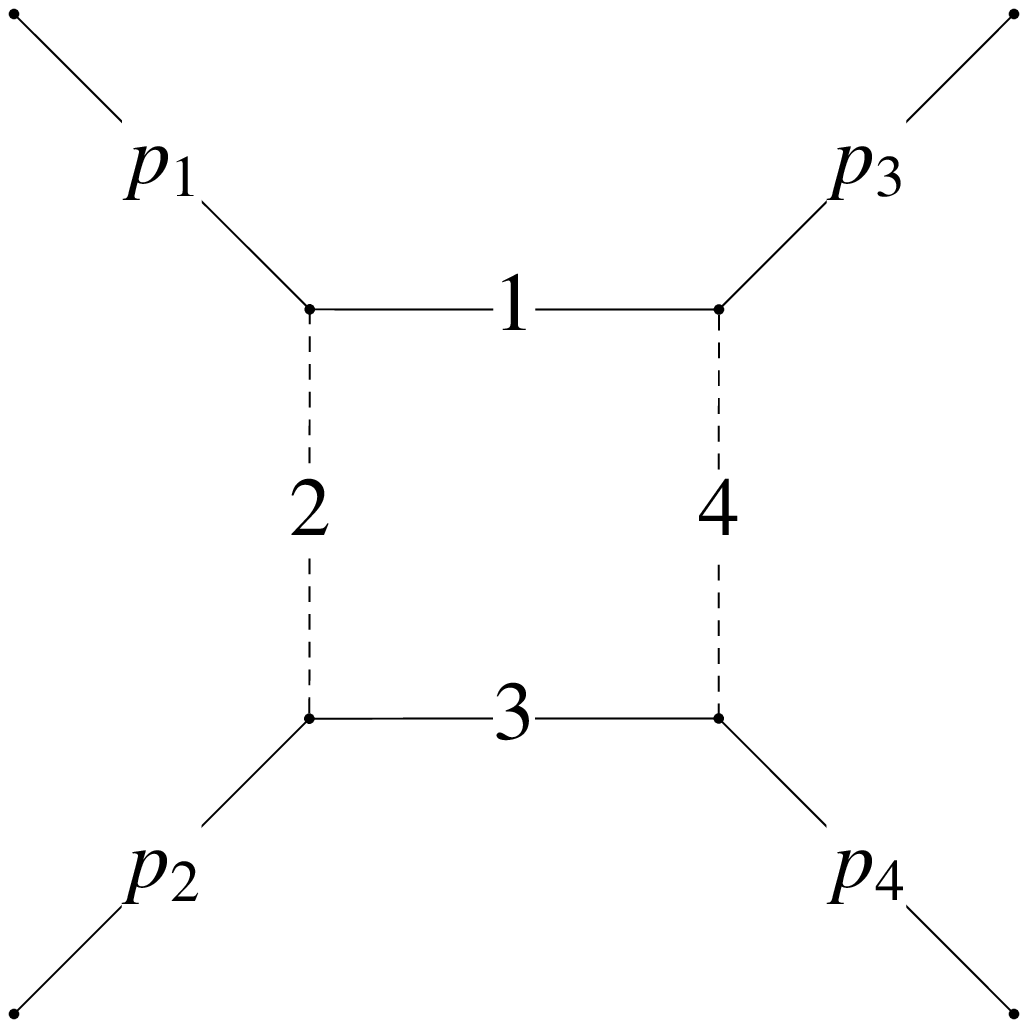}}
\hspace{1cm}
\subfloat[(2a)]{\includegraphics[width=0.3\textwidth]{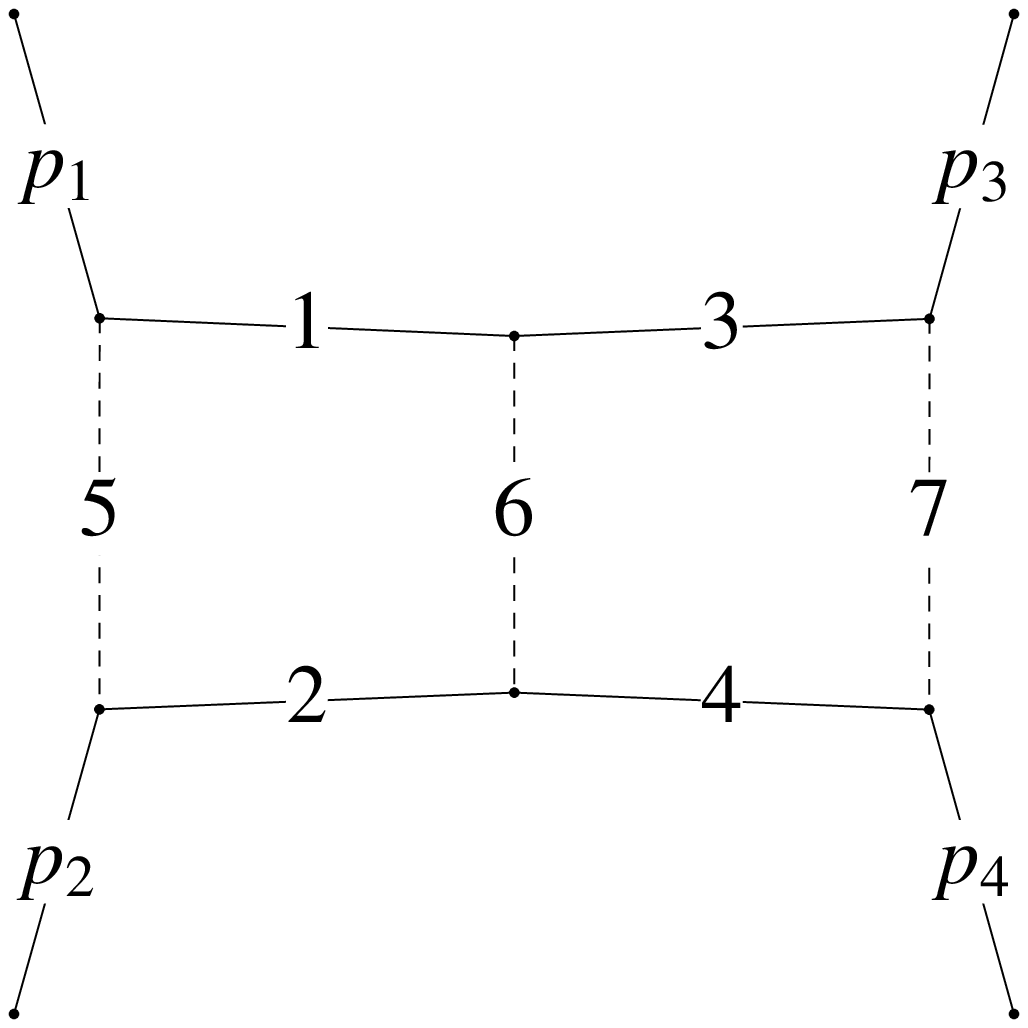}}
\caption{One-loop  and two-loop family of integrals (2a) for Bhabha scattering. 
Solid lines indicate massive propagators, and dashed lines massless ones.}
\label{fig:bhabha2def}
\end{center}
\end{figure}

In this paper we present a solution of this problem for the family of planar diagrams of the first type
contributing to massive two-loop Bhabha scattering -- see  Fig.~\ref{fig:bhabha2def}(2a).
For the other planar and non-planar family, see e.g. Fig. 1 of \cite{Heinrich:2004iq}.
Explicitly, we are dealing with dimensionally regularized integrals
\bea \label{deftwoloop2a}
G_{a_1,a_2,\ldots,a_9}(s,t,m^2;D) &=&
\int\int \frac{\dr^D k_1 \, \dr^D k_2}{(-k_1^2 + m^2)^{a_1}[-(k_1 + p_1 + p_2)^2 + m^2]^{a_2}[-k_2^2 +  m^2]^{a_3}}
\nn \\ && \hspace*{-30mm}
\times \frac{[-(k_2 + p_1)^2]^{-a_8}[-(k_1 - p_3)^2]^{-a_9}}{
[-(k_2 + p_1 + p_2)^2 + m^2]^{a_4}
[-(k_1 + p_1)^2]^{a_5} [-(k_1 - k_2)^2]^{a_6} [-(k_2 - p_3)^2]^{a_7}}
\label{gen2a}
\eea
labelled by seven indices for propagators and two indices for irreducible numerators.
In other words, the $a_{i}$ can take any integer values, with $a_{8}$ and $a_{9}$ 
being zero or negative.
The external legs are on-shell, $p_i^2=m^2$, for $i=1,2,3,4$ and 
the Mandelstam variables are $s=(p_1+p_2)^2$ and $t=(p_1+p_3)^2$, and $u = (p_2+p_3)^2 = -s-t+4 m^2$.
In the following, we will often suppress the explicit dependence of the 
integrals on the kinematical variables.

We will show that one can resolve all issues mentioned above within the DE approach.
The crucial point is to apply a new strategy of solving DE for master integrals which was
suggested \cite{Henn:2013pwa} by one the authors of the present paper and already applied to
to the evaluation of all the planar three-loop four-point massless on-shell master integrals
\cite{Henn:2013tua}.
The key ingredient of this strategy is to turn, after revealing master integrals by
integration-by-parts (IBP) relations \cite{Chetyrkin:1981qh}, to
a much more convenient basis of master integrals consisting of 
{\em pure} functions of uniform {\em weight}.\footnote{In the physics literature, this is
often referred to as uniform degree of ``transcendentality''. We prefer the more concise
notion of weight for iterated integrals.} 
Let us recall that for iterated integrals over logarithmic differential forms, 
i.e. Chen iterated integrals \cite{Chen1997} (see also the recent lecture notes \cite{iterated1,iterated2}),
the weight is defined as the number of integrations. For rational arguments of the logarithms,
these functions are the commonly used Goncharov polylogarithms  \cite{Goncharov:1998kja}, or closely
related hyperlogarithms.
We will say that a linear combination of such 
functions has uniform (i.e. homogeneous) weight if all its summands have the same weight. Finally, a function is 
called pure if only $\mathbb{Q}$-linear 
combinations are allowed. As a consequence given such a pure function, 
the weight of its differential is lowered by one unit. 
This property is essential for such functions to be able to satisfy simple
differential equations.

The first step of the method of DE
is to take derivatives of a given
master integral with respect to kinematical invariants and masses. 
The result of this differentiation is then expressed in terms of
Feynman integrals of the given family and, according to the known IBP
reduction\footnote{In our calculations, we use {\tt FIRE} \cite{Smirnov:2008iw,Smirnov:2013dia}
to perform an IBP reduction.}, in terms of the master integrals.
In this way, one obtains a system of first-order differential equations
for the master integrals, and can then try to solve this system with appropriate boundary 
conditions.  

We find a basis choice that puts the differential equations into the canonical form proposed in ref. \cite{Henn:2013pwa},
\begin{align}\label{diffeqcanonical}
d\, f = \eps\, d \tilde{A} \, f \,.
\end{align}
Here $f(s/m^2,t/m^2;\eps)$ is a list of master integrals for the family defined in eq. (\ref{deftwoloop2a}),
$\tilde{A}$ is a matrix
and $d\,\tilde{A}$  which is a logarithmic one-form, independent of $\eps$.

This implies that the general solution, to all orders in $\eps$ can immediately be written down in terms of 
Chen iterated integrals, where the integration kernels follow from the one-form $d \tilde{A} $.
We will discuss these functions and their relation to functions more commonly used in the physics literature
in more detail in the main text.
Moreover, the arguments of the logarithms in $\tilde{A}$ completely specify the class of functions
needed to describe the Feynman integrals to all orders in the $\eps$ expansion.
At low orders in this expansion, up to weight four which is relevant to NNLO computations,
only a subset of these functions is needed, and we will describe this in details.
Another feature of eq. (\ref{diffeqcanonical}) is that it makes the singularity structure of the integrals
completely transparent. Among other things, this is extremely helpful when determining
the boundary constants of integration. The latter follow from simple physical considerations, 
such as the absence of certain branch cuts in unphysical channels, and are straightforward to implement
in this approach.

Our results show that, contrary to common belief, two-loop integrals involving masses need not be
complicated, if thought about in the right way. 
We present the results up to degree four for all except one integrals in terms of a subset of Goncharov polylogarithms,
which one may call two-dimensional harmonic polylogarithms.
For one integral, and more generally for higher degree, the solution is written in terms of Chen iterated integrals.
We discuss one degree four example in detail and explain how to evaluate this integral numerically.
For the Goncharov polylogarithms, standard tools for numerical evaluation are available~\cite{Bauer:2000cp}.

This paper is organized as follows.
In section \ref{sec:oneloop} we discuss the one-loop case in detail within our method.
The reason is that with small changes the two-loop case
can be treated in the same way. Also, this allows to expand results for one-loop
integrals to any order in $\eps$, if they are needed for future computations.
We discuss in detail several different analytic
representations of the answer, using Chen iterated integrals, 
Goncharov polylogarithms, and multiple polylogarithms.
In section \ref{sec:twoloop}, we apply the same technique to two loops.
We discuss the class of functions needed and give an explicit
analytic solution for all integrals up to transcendental weight four.
For reasons of brevity, we only show explicitly the result for two
of the most complicated integrals, and provide the formulas for the
remaining integrals in electronic form in an ancillary file.
In section \ref{sec:outlook}, we conclude and discuss future directions.
Technical details are delegated to the appendix.

%
%
%
%
%

\section{Master integrals and differential equations at one loop}
\label{sec:oneloop}

Here we discuss the one-loop case as a useful pedagogical example and
as a preparation for the two-loop analysis. As we will see, most of the
analysis can be directly carried over to the two-loop case.
We start by defining the family of one-loop integrals, as well as convenient variables.
We then discuss the choice of integral basis, the expected analytic structure of the 
integrals, and the differential equations they satisfy.
We explain how to solve the latter to all orders in the $\eps$ expansion and
discuss various possibilities for representing the answer in analytic form.

\subsection{Definitions}

We will discuss the family of one-loop integrals displayed in Fig.~\ref{fig:bhabha2def}(1).
In other words, we have
\bea\label{oneloopfamily}
G_{a_1,\ldots,a_4}
=\int\frac{\dr^D k}{[-k^2 + m^2]^{a_1}[ -(k + p_1)^2]^{a_2}[-(k + p_1 + p_2)^2 + m^2 ]^{a_3}[  -(k - p_3)^2]^{a_4}}
\,,
\eea
for any integer values of the $a_{i}$.
We can always make integrals dimensionless by multiplying them by appropriate factors.
Therefore we can parametrize them by following dimensionless variables, which we choose as
\begin{align}\label{defxy}
\frac{-s}{m^2} = \frac{(1-x)^2}{x} \,,\qquad \frac{-t}{m^2} = \frac{(1-y)^2}{y} \,.
\end{align}
The new variables $x$ and $y$ have the advantage of resolving some square roots 
typical for massive scattering processes.
It is useful to define all functions in the Euclidean region $s<0, t<0$, where
the results for planar integrals are manifestly real, and analytically continue those
formulas to the physical region.

We may note that the definition (\ref{defxy}) is invariant under inversions of $x$ and $y$.
Depending on the $x$-dependent normalization factors, all functions will have definite
antisymmetry or symmetry properties under these transformations.
This symmetry also implies that we can take $|x|<1, |y|<1$ without loss of generality.

There are several special values for $x$ and $y$ corresponding to important physical limits
of the integrals.
Focusing on the variable $s$ (or, equivalently, $x$), we have
\begin{align}\label{boundarycases}
x=0 \;\leftrightarrow\; s = \infty\,,\qquad x = 1 \;\leftrightarrow\; s = 0 \,\qquad x=-1\; \leftrightarrow\; s = 4 m^2 \,,
\end{align}
and similarly for $s\leftrightarrow t$.
Physically, the limiting cases of eq. (\ref{boundarycases}) correspond to the Regge limit, the soft limit, and the threshold of
creating a pair of massive particles, respectively.
Moreover, $u=0$ corresponds to $x=-y$.
Other special configurations are related to the above by the $x \to 1/x$ and $y \to 1/y$ symmetry.

%
%
%
%
%

\subsection{Analytic structure of the integrals}
\label{sec:analytic}

As we will see, the integrals of the family (\ref{oneloopfamily}) will give rise
to two-variable generalizations of multi-valued functions similar to polylogarithms.
When writing down the answer for these integrals, it is important to understand
their branch cut structure. This knowledge can also be helpful for determining integration
constants. For instance, the absence of singularities at certain kinematical points can lead
to analytic constraints determining boundary constants.
This will be particularly transparent in the canonical form (\ref{diffeqcanonical}) of the differential equations.

When thinking about the analytic structure of the answer it is useful to have the example
of the one-loop box function in mind, see Fig.~\ref{fig:bhabha2def}(1),
which has the Feynman parametrization
\begin{align}\label{oneloopFeynman}
G_{1,1,1,1} = \Gamma(1+\eps) \,  \int_{0}^{\infty}  \, 
\frac{  \left( \prod_{i=1}^{4} d \alpha_i \right) \delta(1-\sum_{i=1}^{4} \alpha_i )}{ [\alpha_1 \alpha_3 (-s) + \alpha_2 \alpha_4 (-t) + m^2 (\alpha_1 +\alpha_3 )^2-i 0]^{2+\eps}} \,.
\end{align}
The  $-i0$ prescription can be absorbed into $m^2$, or by addition $+i0$ to $s$ and $t$.
From the physical point of view it is clear that $G_{1,1,1,1}$ has branch cuts in the $t$-channel starting at $t=0$ and extending along the positive $t$ axis, corresponding to two massless particles in the $t$-channel unitarity cut. Likewise, one expects a branch cut starting at $s=4 m^2$ (and extending along the positive $s$ axis) from the $s$-channel cut through two massive lines.
One can arrive at the same conclusions by analyzing possible singularities of eq. (\ref{oneloopFeynman}), using Landau equations \cite{Landau:1959fi}.
Moreover, for planar graphs, we expect there to be no branch cuts at $u=0$, where $s+t+u = 4 m^2$, and hence $x=-y$.
Another interesting limit is $s \to 0$, or equivalently $x \to 1$.
Analyzing the above Feynman parametrization (\ref{oneloopFeynman}) 
we expect that this limit does not introduce new divergences (unlike e.g. the $t \to 0$ limit),
and therefore should commute with the $\eps \to 0$ expansion, and give a finite answer.

Finally, it is clear that the answer must be real-valued in the Euclidean region $s<0, t<0$.
As already mentioned, the above information about the analytic structure of the integrals will be extremely useful 
in determining the boundary constants.

The explicit results we present will be written in a form where each term is 
real-valued in the Euclidean region $0<x<1, 0<y<1$. One can extend them to the physical
region by analytic continuation, see e.g. \cite{Gehrmann:2002zr,Bonciani:2003cj} .

%
%
%
%
%

\subsection{Choice of integral basis}

\begin{figure}[t] 
\captionsetup[subfigure]{labelformat=empty}
\begin{center}
\subfloat[(1)]{\includegraphics[width=0.2\textwidth]{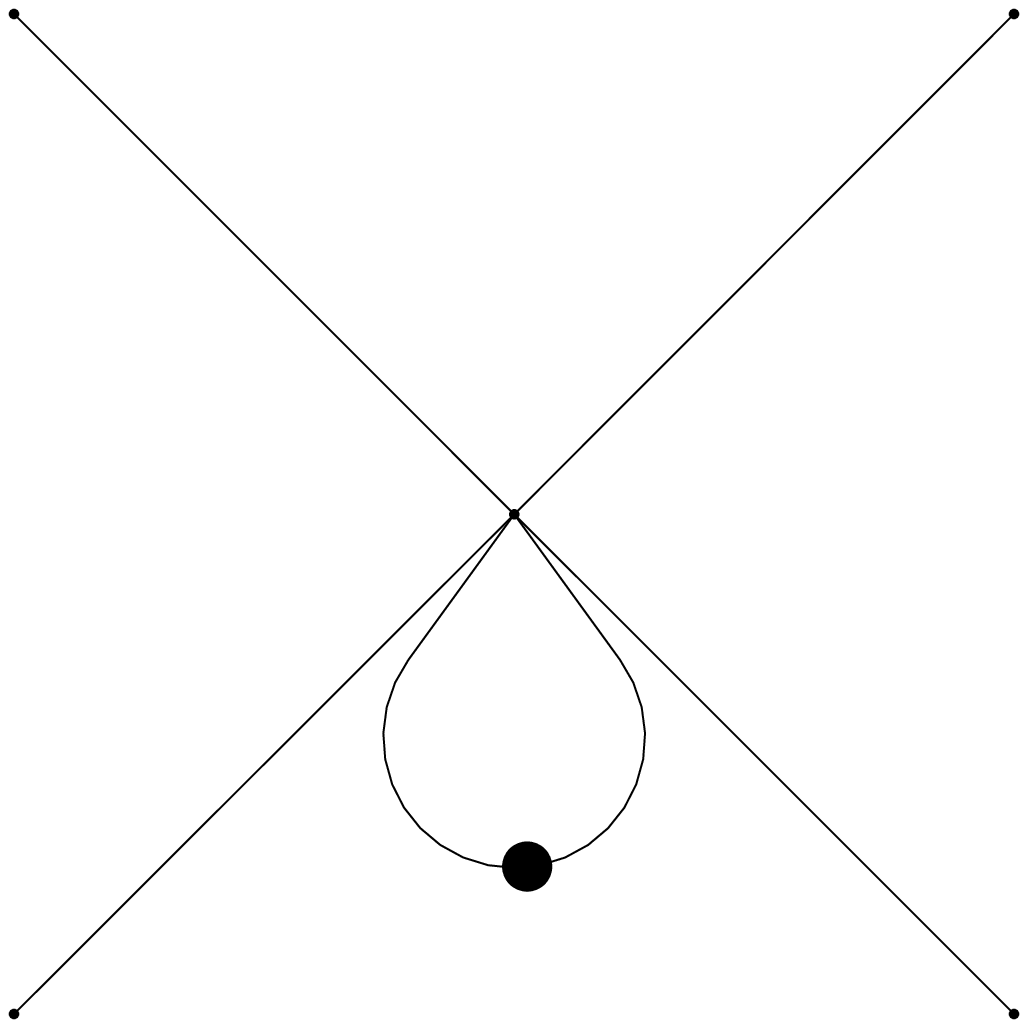}}
\subfloat[(2)]{\includegraphics[width=0.2\textwidth]{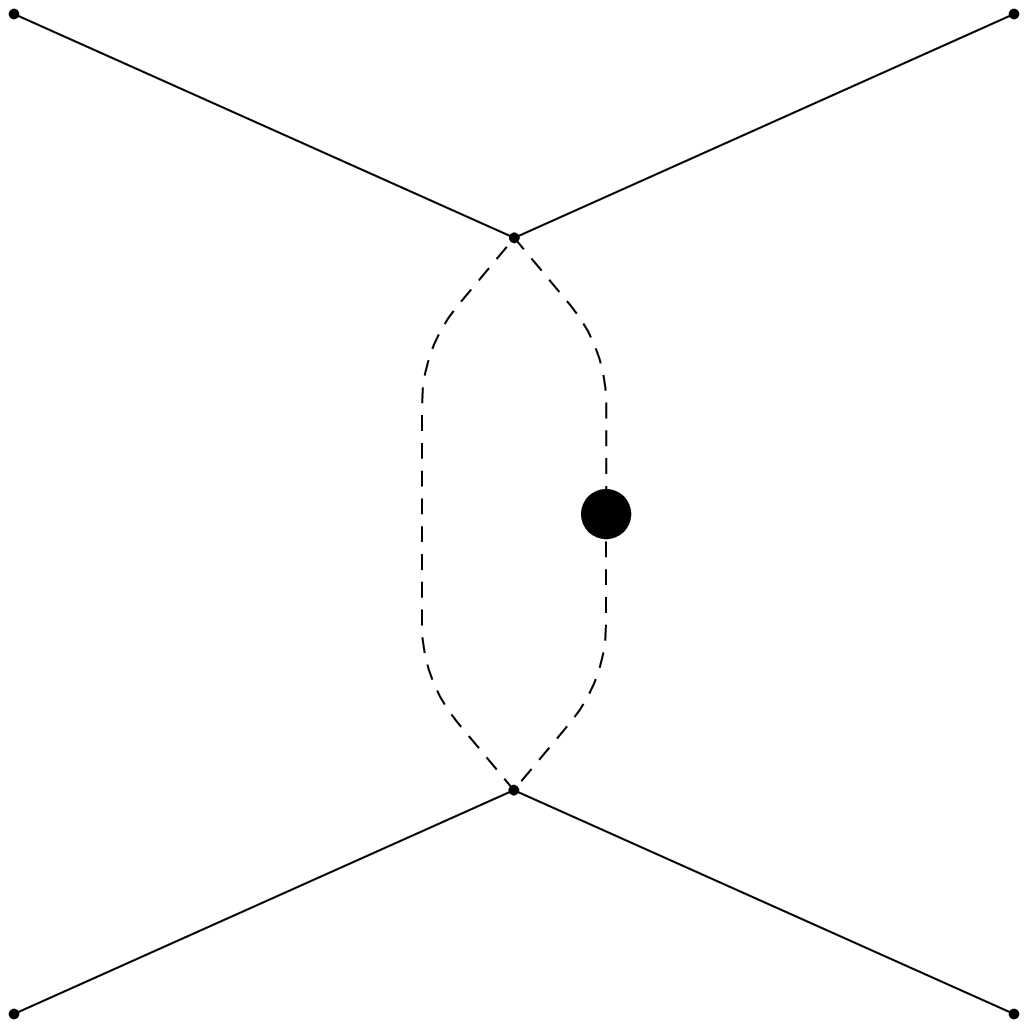}}
\subfloat[(3)]{\includegraphics[width=0.2\textwidth]{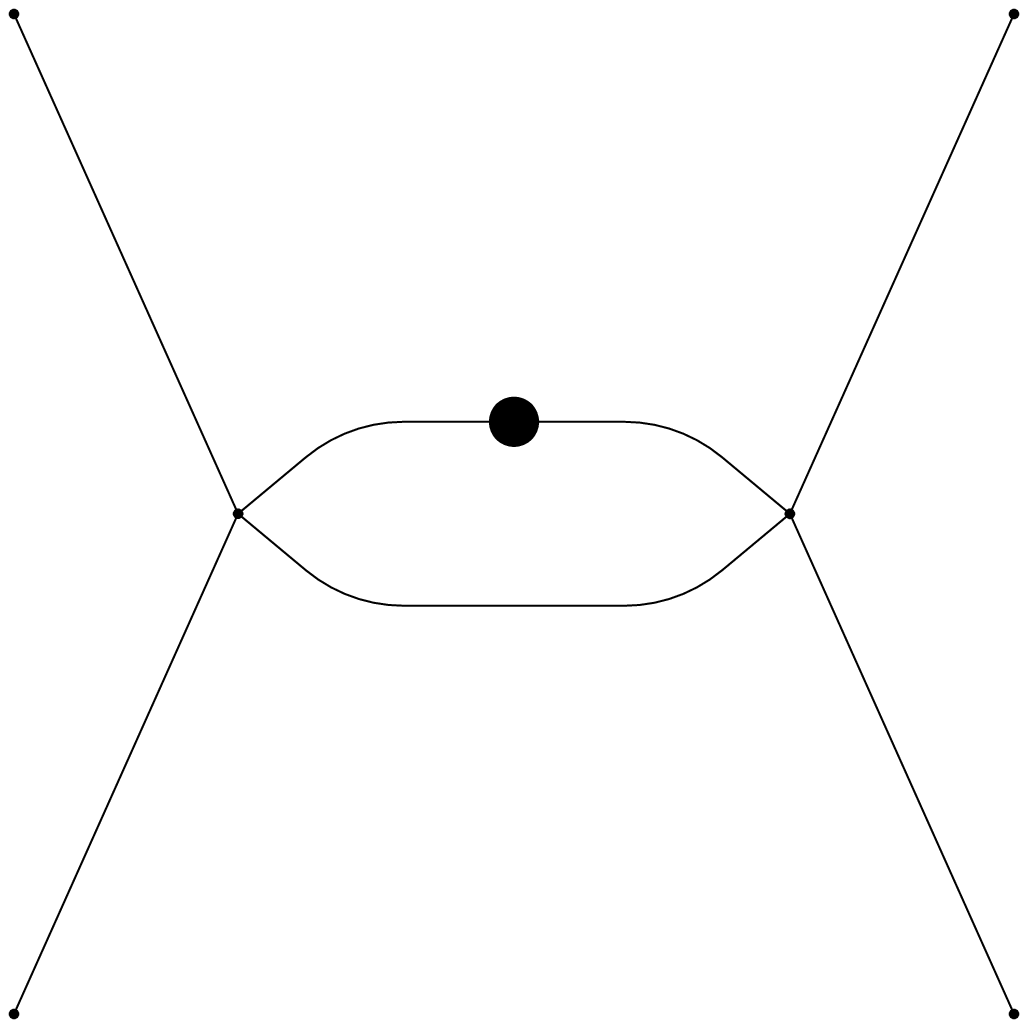}}
\subfloat[(4)]{\includegraphics[width=0.2\textwidth]{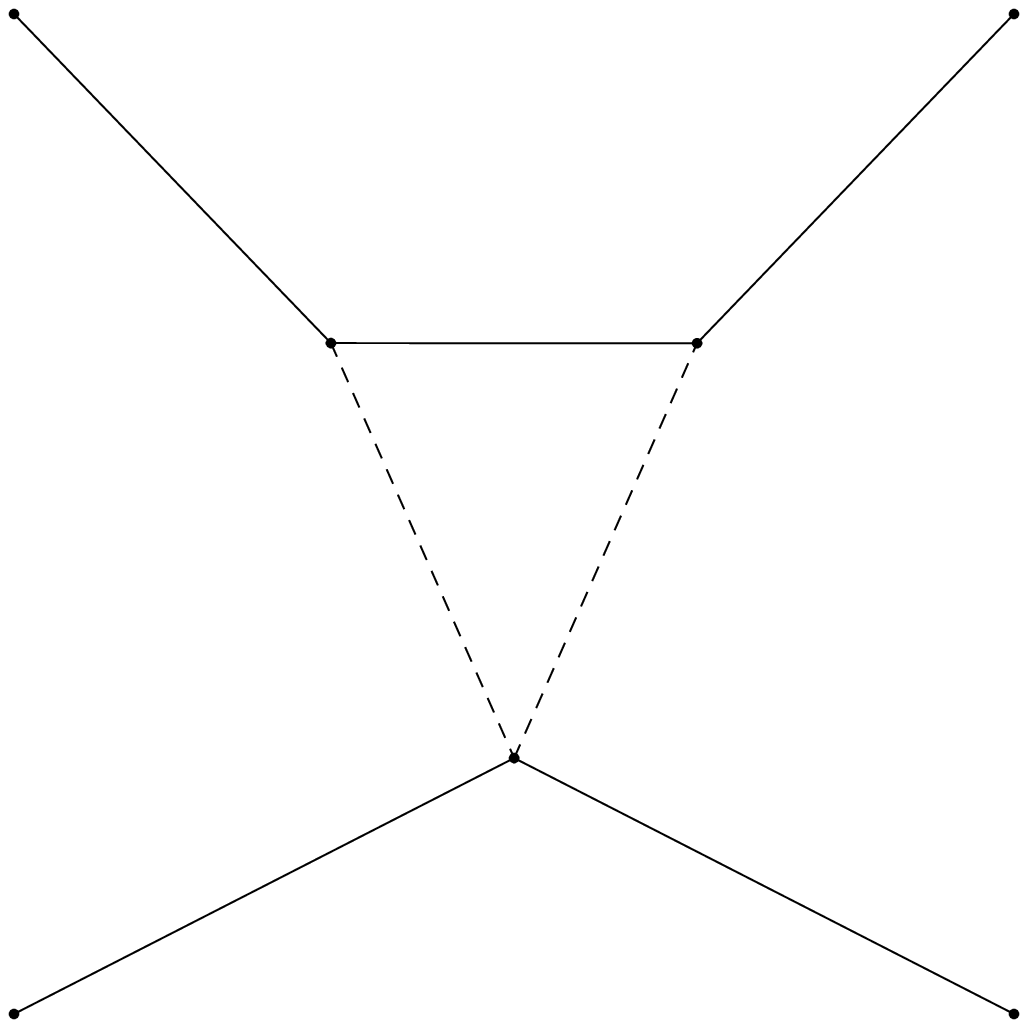}}
\subfloat[(5)]{\includegraphics[width=0.2\textwidth]{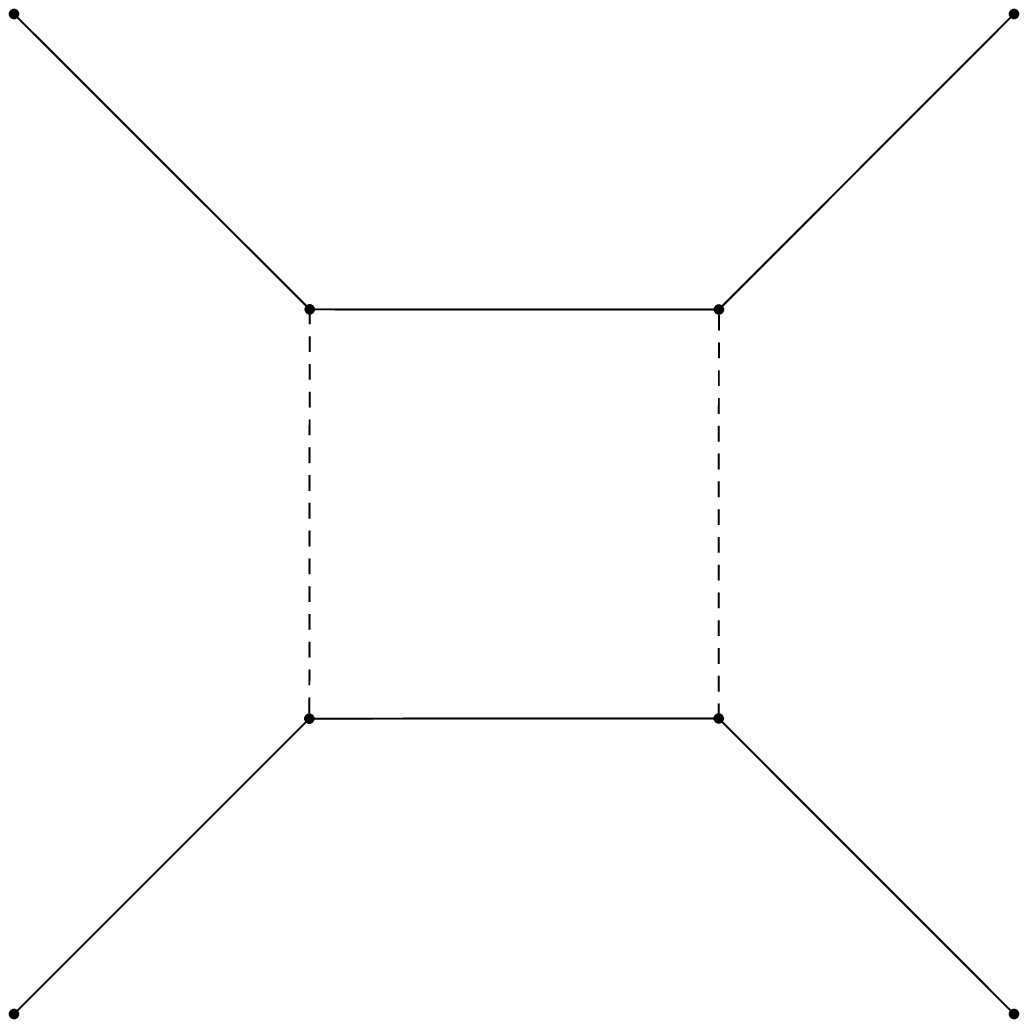}}
\newline
\caption{Master integrals for the integral family of one-loop Bhabha integrals.
}
\label{fig:basisbhabha1}
\end{center}
\end{figure}
We choose candidate integrals that are expected to be pure functions
of uniform weight, as explained in ref. \cite{Henn:2013pwa}, see also
ref. \cite{Henn:2013tua} for examples of the application of the method.
This analysis leads us to the following choice, depicted in Fig.~\ref{fig:basisbhabha1},
\begin{align}\label{Bhabha1basis0}
f_i= (m^2)^\eps  \, e^{\eps \gamma_{\rm E}} \, g_{i} \, 
\end{align}
with
\begin{align}\label{Bhabha1basis}
g_1 =& \eps  \, G_{2,0,0,0}  \,,\\
g_2 =& \eps  t \, G_{0,2,0,1}  \,,\\
g_3 =& \eps \,  \sqrt{(-s)(4 m^2-s)} \, G_{2,0,1,0}  \,,\\
g_4 =& -2 \eps^2 \,  (4 m^2-t)(-t)\, G_{1,1,0,1}  \,,\\
g_5 =& -2 \eps^2 \, \sqrt{(-s)(4 m^2-s)}  t \, G_{1,1,1,1}  \,.
\end{align}
We normalized all integrals such that they have a Taylor expansion around $\eps=0$, i.e.
\begin{align}
f_{i} = \sum_{k \ge 0} \eps^k \, f^{(k)}_{i} \,.
\end{align}
Of course, the tadpole and bubble integrals are elementary (see e.g. \cite{Smirnov:2012gma}),
\begin{align}
f_1 =& \eps\, \Gamma(\eps)\, e^{\eps \gamma_{\rm E}} \,, \label{elementary1}\\
f_2 =& -\eps \, \Gamma(1 - \eps) \Gamma(-\eps) \Gamma(1 + \eps)/\Gamma(1-2\eps) 
\left(\frac{y}{(1-y)^2}\right)^\eps e^{\eps \gamma_{\rm E}}\,. \label{elementary2}
\end{align}
They will provide simple boundary conditions to all orders in $\eps$.
One may observe that they have uniform weight in their expansion in $\eps$.

Let us now analyze the system of differential equations that these integrals satisfy.

%
%
%
%
%

\subsection{One loop differential equations and integration}
With the choice of integral basis given in eq. (\ref{Bhabha1basis}), 
we find the following system of differential equations.
We have eq. (\ref{diffeqcanonical}), with the matrix $\tilde{A}$ given by
\begin{align}\label{diffeqBhabha1}
\tilde{A} = \Big[ &
{\scriptsize\left(
\begin{array}{ccccc}
 0 & 0 & 0 & 0 & 0 \\
 0 & 0 & 0 & 0 & 0 \\
 -1 & 0 & 1 & 0 & 0 \\
 0 & 0 & 0 & 0 & 0 \\
 0 & 4 & 0 & 0 & 0
\end{array}
\right)} \, 
\log x + 
{\scriptsize\left(
\begin{array}{ccccc}
 0 & 0 & 0 & 0 & 0 \\
 0 & 0 & 0 & 0 & 0 \\
 0 & 0 & -2 & 0 & 0 \\
 0 & 0 & 0 & 0 & 0 \\
 0 & 0 & -8 & 0 & -2
\end{array}
\right)}
\, \log(1+x) + 
{\scriptsize\left(
\begin{array}{ccccc}
 0 & 0 & 0 & 0 & 0 \\
 0 & 1 & 0 & 0 & 0 \\
 0 & 0 & 0 & 0 & 0 \\
 2 & -2 & 0 & 0 & 0 \\
 0 & 0 & -4 & 0 & 0
\end{array}
\right)}
\, 
\log y + \nonumber \\
& + 
{\scriptsize\left(
\begin{array}{ccccc}
 0 & 0 & 0 & 0 & 0 \\
 0 & 0 & 0 & 0 & 0 \\
 0 & 0 & 0 & 0 & 0 \\
 0 & 0 & 0 & 2 & 0 \\
 0 & 0 & 0 & 0 & 0
\end{array}
\right)}
\, \log(1+y)
 + 
{\scriptsize \left(
\begin{array}{ccccc}
 0 & 0 & 0 & 0 & 0 \\
 0 & -2 & 0 & 0 & 0 \\
 0 & 0 & 0 & 0 & 0 \\
 0 & 0 & 0 & -2 & 0 \\
 0 & 0 & 0 & 0 & -2
\end{array}
\right)}
\, \log(1-y)
 + \nonumber \\
& + 
{\scriptsize\left(
\begin{array}{ccccc}
 0 & 0 & 0 & 0 & 0 \\
 0 & 0 & 0 & 0 & 0 \\
 0 & 0 & 0 & 0 & 0 \\
 0 & 0 & 0 & 0 & 0 \\
 0 & 0 & 4 & 2 & 1
\end{array}
\right)}
\, \log(x+y)
 + 
{\scriptsize \left(
\begin{array}{ccccc}
 0 & 0 & 0 & 0 & 0 \\
 0 & 0 & 0 & 0 & 0 \\
 0 & 0 & 0 & 0 & 0 \\
 0 & 0 & 0 & 0 & 0 \\
 0 & 0 & 4 & -2 & 1
\end{array}
\right)}
\, \log(1+x y)
\Big]  \,.
\end{align}
Let us analyze the logarithmic differential forms $d \tilde{A}$ appearing in the differential equation (\ref{diffeqcanonical}).
One may note that there is no $d\, \log(1-x)$ term, which is related to the boundary condition at $x=1$ (cf. the discussion in section \ref{sec:analytic}.) 
Note however, that the finiteness of the functions $f_{i}$ at $x=1$ does not imply the absence of such a term in the differential equations,
as we will see in the two-loop analysis. Therefore, for more generality, let us allow for such a term in our discussion.

As was pointed out in ref. \cite{Henn:2013pwa}, a differential equation of the type  (\ref{diffeqcanonical}) 
makes it manifest that the solution is simple.
The latter can always be written down in terms of Chen iterated integrals that have manifest transcendental weight.
Moreover, experience shows that the boundary conditions for planar integrals usually can be obtained 
from simple physical conditions.
Indeed, we can demand that the solution have the properties discussed in section \ref{sec:analytic}.
It can be seen that these conditions, together with the elementary bubble and tadpole integrals of eqs. (\ref{elementary1}) and (\ref{elementary2}),
determine all boundary constants.
Therefore one can claim that the problem is essentially solved when a form (\ref{diffeqcanonical}) of the
differential equations is found.

To some extent the choice of how to present the explicit solution, or what is meant by explicit, is a matter
of taste and depends also on the applications one has in mind.
Let us elaborate on the various possibilities in the following subsections.

\subsection{Symbol of the solution}

Before writing down explicit formulas for the solution,
let us mention that as a corollary of the differential equations (\ref{diffeqcanonical}), 
the symbol 
\cite{arXiv:math/0606419,2009arXiv0908.2238G} 
of the answer is immediately determined.  The reader unfamiliar with symbols
may think of the symbol as encoding the differential form 
$d \tilde{A}$
needed to define
the functions through iterated integrals, as discussed in the next section.
In the present case, the symbol alphabet is seen to be
\begin{align}\label{symbolalphabet1}
\{ \, x\,, 1\pm x \,, y \,, 1\pm y \,, x+y \,, 1+x y  \,\}
\end{align}
where we again included $1-x$ in view of the two-loop case, although it was not needed at one loop.

The boundary conditions discussed in section \ref{sec:analytic} and coming from the elementary
integrals of eqs.  (\ref{elementary1}) and (\ref{elementary2}) 
give us
\begin{align}\label{sol0}
f^{(0)} = \{ 1,1,0,0,0 \}\,.
\end{align}
This completely specifies the symbol.
It is simply obtained by iterations of the matrix $\tilde{A}$.
Let us take the box integral $f_{5}$ of eq. (\ref{Bhabha1basis0})  as a specific example.
Using the notation $[\ldots]$ for a symbol, e.g.
$\mathcal{ S} ( {\rm Li}_{2}(x) ) = - [1-x,x]$,
one obtains, in the first
few orders of the $\eps$ expansion,
\begin{align}
\mathcal{S}(f_{5}^{(0)}) =& \, 0 \,,  \\
\mathcal{S}(f_{5}^{(1)}) =& \,4 [x]  \,,\\
\mathcal{S}(f_{5}^{(2)}) =& \,-8 [x, 1 - y] + 4 [x, y] - 8 [1 - y, x] + 4 [y, x]  \,,\\
\mathcal{S}(f_{5}^{(3)}) =& \, 8 [x, x, 1 + x] + 4 [x, x, y] - 4 [x, x, x + y] - 
 4 [x, x, 1 + x y] \nonumber \\ & - 16 [x, 1 + x, 1 + x] - 8 [x, 1 + x, y] + 
 8 [x, 1 + x, x + y] \nonumber \\ &  + 8 [x, 1 + x, 1 + x y] + 
 16 [x, 1 - y, 1 + x] + 16 [x, 1 - y, 1 - y] \nonumber \\ &  - 
 8 [x, 1 - y, x + y] - 8 [x, 1 - y, 1 + x y] - 
 8 [x, y, 1 + x] \nonumber \\ &  - 8 [x, y, 1 - y] + 4 [x, y, x + y] + 
 4 [x, y, 1 + x y] + 16 [1 - y, x, 1 + x] \nonumber \\ &  + 
 16 [1 - y, x, 1 - y] - 8 [1 - y, x, x + y] - 
 8 [1 - y, x, 1 + x y] \nonumber \\ & + 16 [1 - y, 1 - y, x] - 
 8 [1 - y, y, x]  + 8 [1 - y, y, x + y]  \nonumber \\ & - 
 8 [1 - y, y, 1 + x y] - 8 [y, x, 1 + x] - 8 [y, x, 1 - y]  \nonumber \\ &+ 
 4 [y, x, x + y] + 4 [y, x, 1 + x y] - 8 [y, 1 - y, x] x+ 
 4 [y, y, x] \nonumber \\ & -  4 [y, y, x + y] + 4 [y, y, 1 + x y]  \,.
\end{align}
This can obviously be continued to any desired order. This
is very useful because it specifies exactly which classes of integral
functions are needed to a given order in $\eps$.

The attentive reader might be surprised about the previous sentence, since 
equation (\ref{diffeqcanonical}), together with the explicit logarithmic 
form of $\tilde{A}$ in (\ref{diffeqBhabha1}), already specify the function
class required to all orders in $\eps$. However, it frequently happens
that the first few orders of the $\eps$ expansion of the master integrals
involve only a subset of such functions, as a consequence of the
specific start of the symbol iteration, see eq. (\ref{sol0}). (Recall that the
latter follows from the boundary conditions.)
We will see an explicit example of this at two loops, where it turns out
that all except one functions at weight four require a smaller function
class compared to the all-order solution in $\eps$.

\subsection{General solution in terms of Chen iterated integrals}
\label{sec:chen}

 In general, an iterative solution in $\eps$ of an equation of the type of 
eq. (\ref{diffeqcanonical}) can be written in terms of Chen iterated integrals. 
We have
\begin{align}\label{path_ordered}
f(x,y,\eps) = \, {\mathbb P}  \, e^{ \eps \, \int_{\mathcal_{C}} d \tilde{A} }\, g(\eps) \,, 
\end{align}
where $g(\eps)$ is constant w.r.t. $x$ and $y$, and represents the boundary conditions. 
$\mathcal{C}$ is a contour in the (in general complex) space of kinematical variables 
connecting a chosen base point $(x_0, y_0)$ to $(x,y)$.
$ {\mathbb P} $ stands for path ordering of the matrix exponential.
We see that this defines $f$ at each order $\eps^n$ as an $n$-fold integral
of an $n$-form. The latter is determined from the matrix $\tilde{A}$.
Note that the integrals are homotopy invariant, as long as $\mathcal{C}$ stays away from
the singularities of $d \, \tilde{A}$.
Moreover, the monodromy properties of the answer are completely transparent,
as they are encoded into $ \tilde{A}$. (See \cite{iterated1,iterated2} for more details.)
As we already mentioned, the boundary conditions can be obtained from physical considerations.

\begin{figure}[t] 
\captionsetup[subfigure]{labelformat=empty}
\begin{center}
{\includegraphics[width=0.3\textwidth]{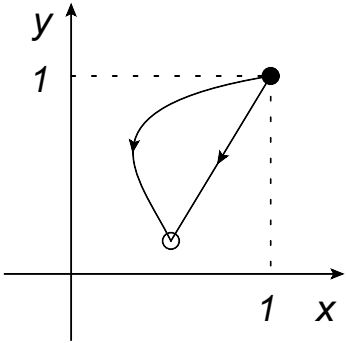}}
\caption{Two integration paths are shown connecting the base point (black dot) and the argument of the function (circle).
The Euclidean region for real $x,y$ with $0<x<1, 0<y<1$ is delimited by dashed lines. Note that this picture is a $\mathbb{R}^2$ slice of $\mathbb{C}^2$. Chen iterated integrals are homotopy invariant 
(on the space where the singular points are removed.) }
\label{fig:paths}
\end{center}
\end{figure}

Some choices of contours $\mathcal{C}$ are illustrated in Fig.~\ref{fig:paths} for real $x,y$ in the
Euclidean region $1>x>0,1>y>0$. However, one should keep in mind that the full picture
should be in $\mathbb{C}^2$.
A concrete application of this will be discussed in section \ref{int11}.

From a mathematical point of view, 
eq. (\ref{path_ordered}) completely solves the problem of computing the master integrals in analytic form. 
This way of writing the solution is also very natural for analytic continuation. 
It is analogous to the definition of the logarithm for complex argument.
Choosing specific integration contours $\mathcal{C}$ one often recovers function classes
more frequently used in the physics literature, as we discuss presently.
We should caution the reader that these other representations, while useful in various regards,
often obscure properties that are manifest in the form (\ref{path_ordered}), and suffer from the
fact that they can be represented in many different but equivalent ways, essentially as a consequence
of the homotopy invariance of eq. (\ref{path_ordered}).

\subsection{Solution in terms of Goncharov polylogarithms}
\label{solGoncharov}

Let us discuss relations of the representation of the general solution of the previous subsection to
functions more frequently used in physics.
We may notice that the arguments of logarithms in eq. (\ref{diffeqBhabha1}) are 
linear in the variables, by virtue of the parametrization (\ref{defxy}). (The same is not
the case when using $-s/m^2$ and $-t/m^2$ as independent variables.) This fact
allows us to immediately rewrite the general solution discussed above in terms
of Goncharov polylogarithms.
The latter are defined as follows,
\begin{align}
G(a_1,\ldots a_n ; z) = \int_0^z \frac{dt}{t-a_{1}} G(a_{2}, \ldots ,a_{n}; t) \,,
\end{align}
with 
\begin{align}
G(a_1 ;z) = \int_0^z \frac{dt}{t-a_{1}}  \,, \qquad a_{1} \neq 0\,.
\end{align}
For $a_{1}=0$, we have $G(\vec{0}_{n};x) = 1/n! \log^n(x)$.
The total differential of a general Goncharov polylogarithm is 
\begin{align}\label{differentialG}
d G(a_1 , \ldots a_{n};z) =&\; G(\hat{a}_{1}, a_{2}, \ldots a_{n}; z) \, d \log\frac{z-a_1}{a_1 - a_2} \nonumber \\
& + G(a_{1}, \hat{a}_{2}, a_{3} , \ldots, a_{n}; z) \,  d \log \frac{a_1 - a_2 }{a_2 - a_3} + \ldots +  \nonumber \\
& + G(a_{1}, \ldots, a_{n-1}, \hat{a}_{n}; z) \, d \log \frac{a_{n-1}-a_{n}}{a_{n}} \,,
\end{align}
where $\hat{a}$ means that this element is omitted.
As discussed in subsection \ref{sec:chen}, one can choose various integration contours $\mathcal{C}$
in the solution (\ref{path_ordered}). It is clear that upon choosing some linear parametrization one can
represent all terms in eq.  (\ref{path_ordered}) as Goncharov polylogarithms with certain parameters $a_{i}$.
Note that different choices of contour can give representations of the same iterated integral
in terms of different sets of Goncharov polylogarithms. In this sense, the formulation in terms
of Chen iterated integrals is more universal.

In the specific case at hand, we found it useful to follow an even more pedestrian approach, which
we found to give a convient form of the answer, as we discuss presently.
At degree one, we obviously have the set of  functions 
\begin{align}
g^{(1)} = \{ \log x \,, \log(1\pm x) \,, \log y \,, \log(1\pm y) \,, \log(1+x y) \,, \log(x + y) \} \,. 
\end{align}
Now, consider a degree two function $f(x,y)$. We can consider the $\partial_y f(x,y)$  derivative.
Its general form is dictated by the symbol alphabet.
Integrating back, we find
\begin{align}
 f(x,y) = \int^y_0\, c_{i,j}\, \sum_{i,j} (\partial_y  \, g^{(1)}_i) g^{(1)}_j +h(x) \,.
\end{align}
For $j=1,2,3$ we have a factorized integral, so we exclude these cases, without loss of generality.
In all other cases, we immediately see that we have the definition of a Goncharov
polylogarithm, with $z=y$, and indices $a_{j}$ drawn from the set $\{0,\pm1, -x,-1/x\}$.
Moreover, $h(x)$ should be given by harmonic polylogarithms (HPL) \cite{Remiddi:1999ew} (cf. Appendix \ref{appendix_hpl}.)
It is easy to see that at higher orders, we will at most have $\mathbb{Q}$-linear combinations of products of the above types of functions.
(Of course, one can derive an equivalent representation by focusing on the $\partial_x$ derivatives.)

Note that this yields a representation that is manifestly real-valued for $0<x<1,0<y<1$, i.e. in the Euclidean region.
This is easy to see: thanks to $0<y<1$, the integration variables $t$ satisfy $0<t<1$, and, given the set of allowed $a_{i}$ relevant
here, one always had $t-a_{i}>0$.

The set of functions is slightly more general than the 2dHPLs introduced in \cite{Gehrmann:2000zt,Gehrmann:2001ck}. 
The reason is that in these references, the index $-1$ was not required (also, w.r.t. their notation we have $x\to -x, y\to -y$.). 
However, this is a very natural generalization, just 
like in the case of HPLs, and it seems natural to continue using the name 
2dHPLs for this slightly larger class of functions.
The fact that these special functions appear in different problems makes them 
even more useful.

For example, here is our result for $f_{5}$ up to degree three for illustration.
\begin{align}\label{examplef5function}
f_{5} =& \eps \Big[ 4 H_0(x) \Big] +   \eps^2 \Big[ 4 G_0(y) H_0(x)-8 G_1(y) H_0(x) \Big] \nonumber \\
&+  \eps^3 \Big[ 
-8 G_0(y) H_{-1,0}(x)+4 G_0(y) H_{0,0}(x)-8 H_0(x) G_{1,0}(y)+16 H_0(x) G_{1,1}(y)
\nonumber \\
& 
+4
   H_0(x) G_{-\frac{1}{x},0}(y)-8 H_0(x) G_{-\frac{1}{x},1}(y)+4 H_0(x) G_{-x,0}(y)-8
   H_0(x) G_{-x,1}(y)
      \nonumber \\
&+8 H_{-1,0}(x) G_{-\frac{1}{x}}(y)
+8 H_{-1,0}(x) G_{-x}(y)-4
   H_{0,0}(x) G_{-\frac{1}{x}}(y)-4 H_{0,0}(x) G_{-x}(y)
      \nonumber \\
&
+4 G_{-\frac{1}{x},0,0}(y)-8
   G_{-\frac{1}{x},0,1}(y)-4 G_{-x,0,0}(y)+8 G_{-x,0,1}(y)+8 H_{-2,0}(x)
      \nonumber \\
&-16
   H_{-1,-1,0}(x)+8 H_{-1,0,0}(x)-4 H_{0,0,0}(x)+\frac{10}{3} \pi ^2
   G_{-\frac{1}{x}}(y)-2 \pi ^2 G_{-x}(y)
      \nonumber \\
&
-\frac{2}{3} \pi ^2 G_0(y)-\frac{4}{3} \pi ^2
   H_{-1}(x)-\frac{7}{3} \pi ^2 H_0(x)+8 \zeta_3
\Big] + \cO(\eps^4)\,.
\end{align}
We note that such formulas lend themselves well to numerical evaluation \cite{Bauer:2000cp}.
We therefore expect that this form of the answer should be sufficient for physical applications.
Nevertheless, before closing this discussion, we wish to comment on another possibility of presenting
the answer in yet another form.

\subsection{Solution to a given weight in terms of multiple polylogarithms}
\label{solmultiple}

The attentive reader will have noticed that some of the Goncharov polylogarithms in 
eq. (\ref{examplef5function}) are just harmonic polylogarithms, up to signs.
More generally, when writing out the solution to a given order in $\eps$, one may try to rewrite
the answer in terms of a minimal number of functions. For example, to degree four, one can write the
answers in terms of classical polylogarithms and one additional function ${\rm Li}_{2,2}$, of certain
arguments. Such a rewriting is not difficult, especially with the aid of symbol analysis.
Examples can be found in many places in the literature, see e.g. \cite{Bonciani:2003cj}.
A simple example at weight two is
\begin{align}
G(-1/x,0;y) = \log y \log(1+x y) + {\rm Li}_{2}\left( - x y \right) \,.
\end{align}
We will not perform such a rewriting at fixed order in $\eps$ here since it  would distract 
from the general pattern of the solution. We do remark however that when physical quantities are
computed, additional simplifications may occur, e.g. due to symmetries or other properties,
and then such an analysis may be more worthwhile.

In the present case, such a rewriting is relatively straightforward for all except one functions at two loops up to weight four.
We believe however that the formulation in terms of Goncharov polylogarithms, or more precisely  the subset of 2dHPLs, of section
\ref{solGoncharov} is more useful in practice. See also the related discussion in section \ref{sec:outlook}.

%
%
%
%
%

\section{Master integrals and differential equations at two loops}
\label{sec:twoloop}
 
Having discussed the one-loop case in great detail in the previous section, we will
find the two-loop analysis relatively straightforward.

 \subsection{Choice of master integrals}
  
\begin{figure}[h] 
\captionsetup[subfigure]{labelformat=empty}
\begin{center}
\subfloat[(1)]{\includegraphics[width=0.2\textwidth]{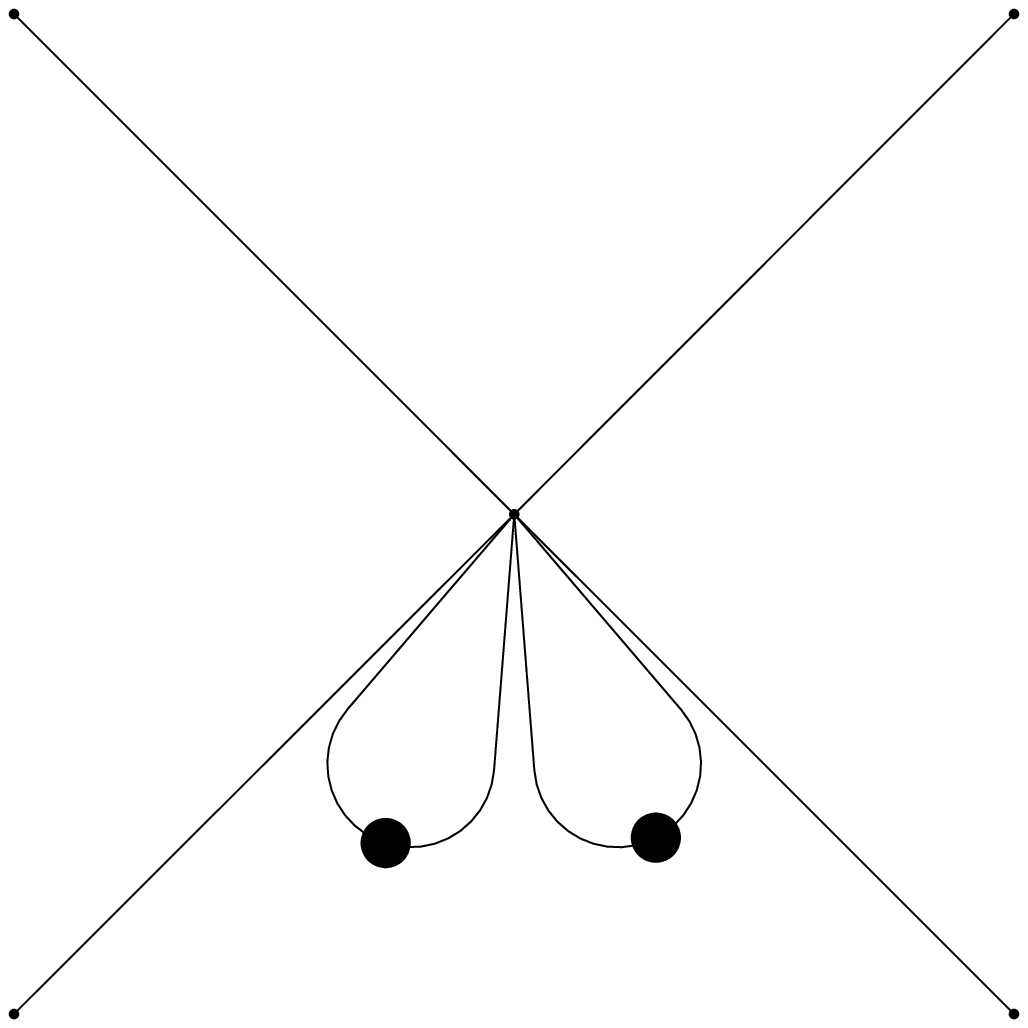}}
\subfloat[(2)]{\includegraphics[width=0.2\textwidth]{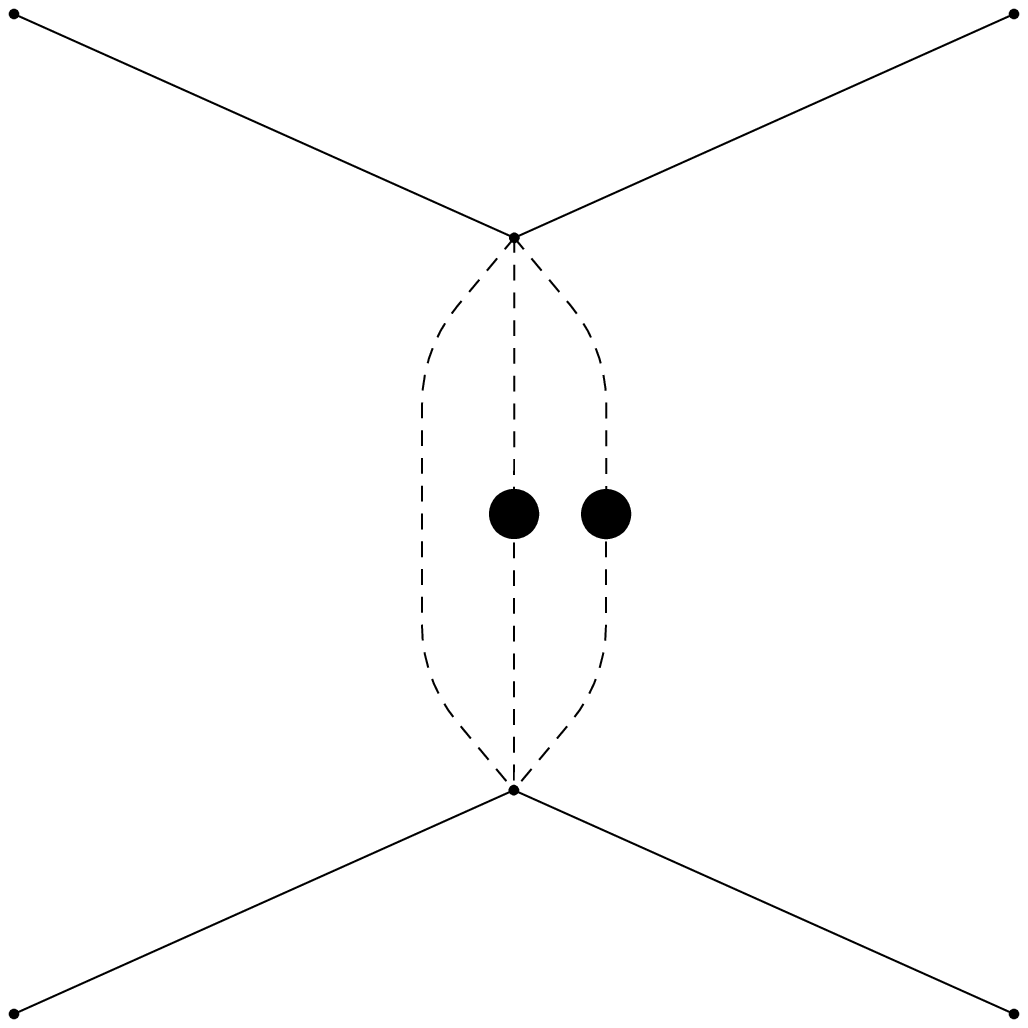}}
\subfloat[(3)]{\includegraphics[width=0.2\textwidth]{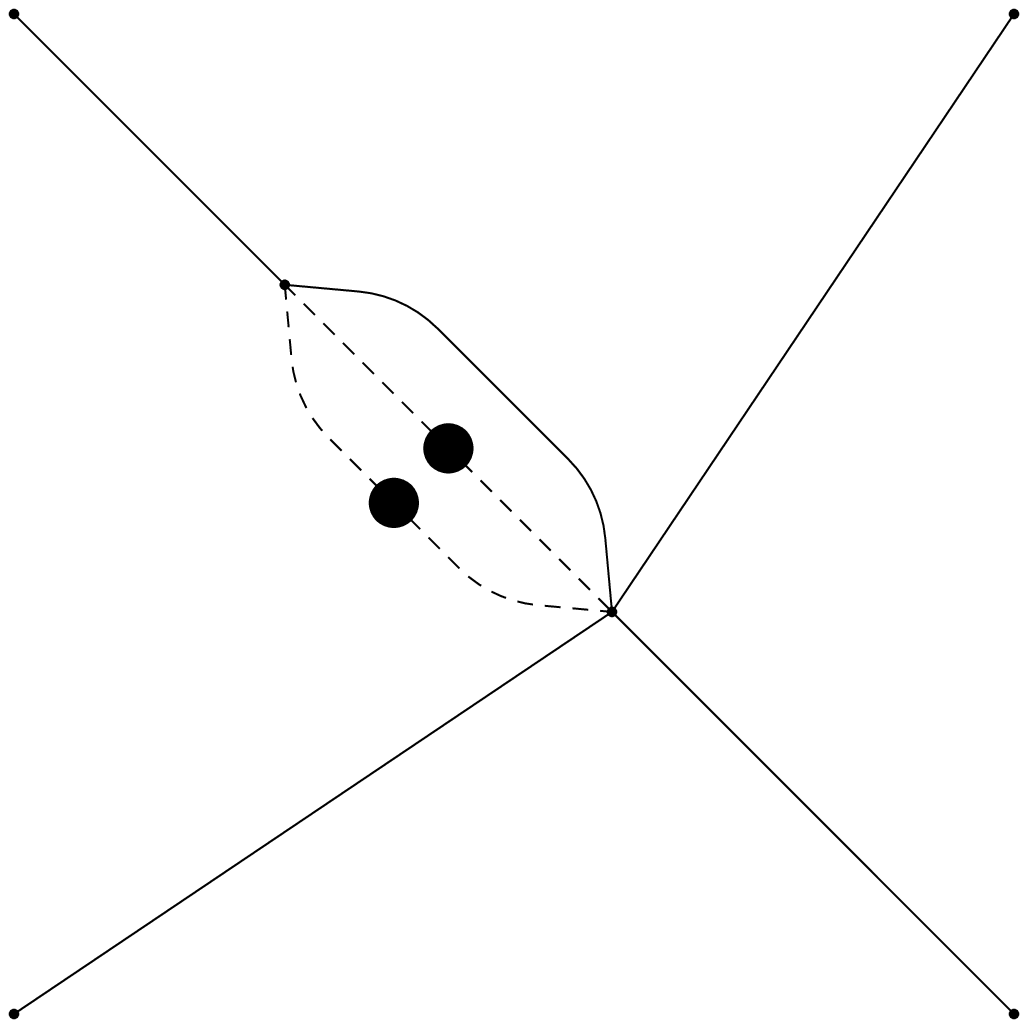}}
\subfloat[(4)]{\includegraphics[width=0.2\textwidth]{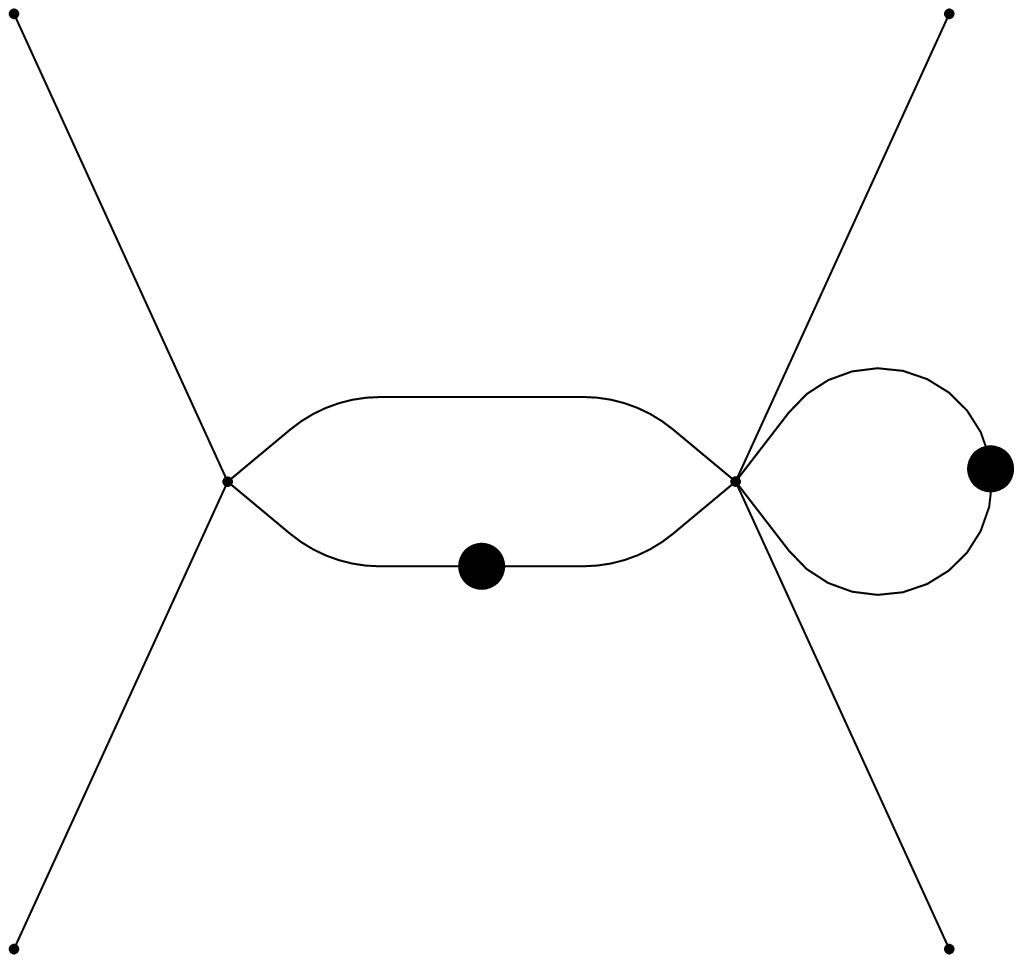}}
\subfloat[(5)${}^{\dagger}$]{\includegraphics[width=0.2\textwidth]{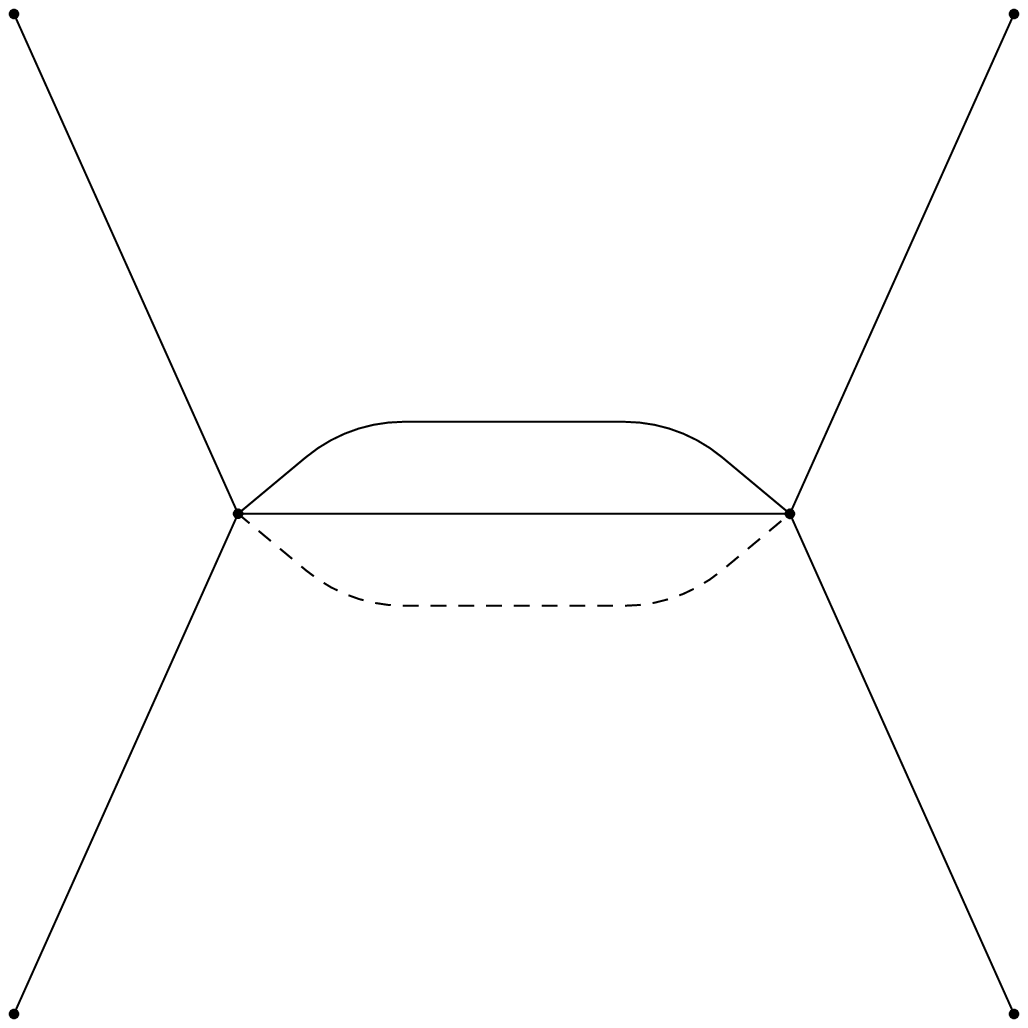}}
\newline
\subfloat[(6)]{\includegraphics[width=0.2\textwidth]{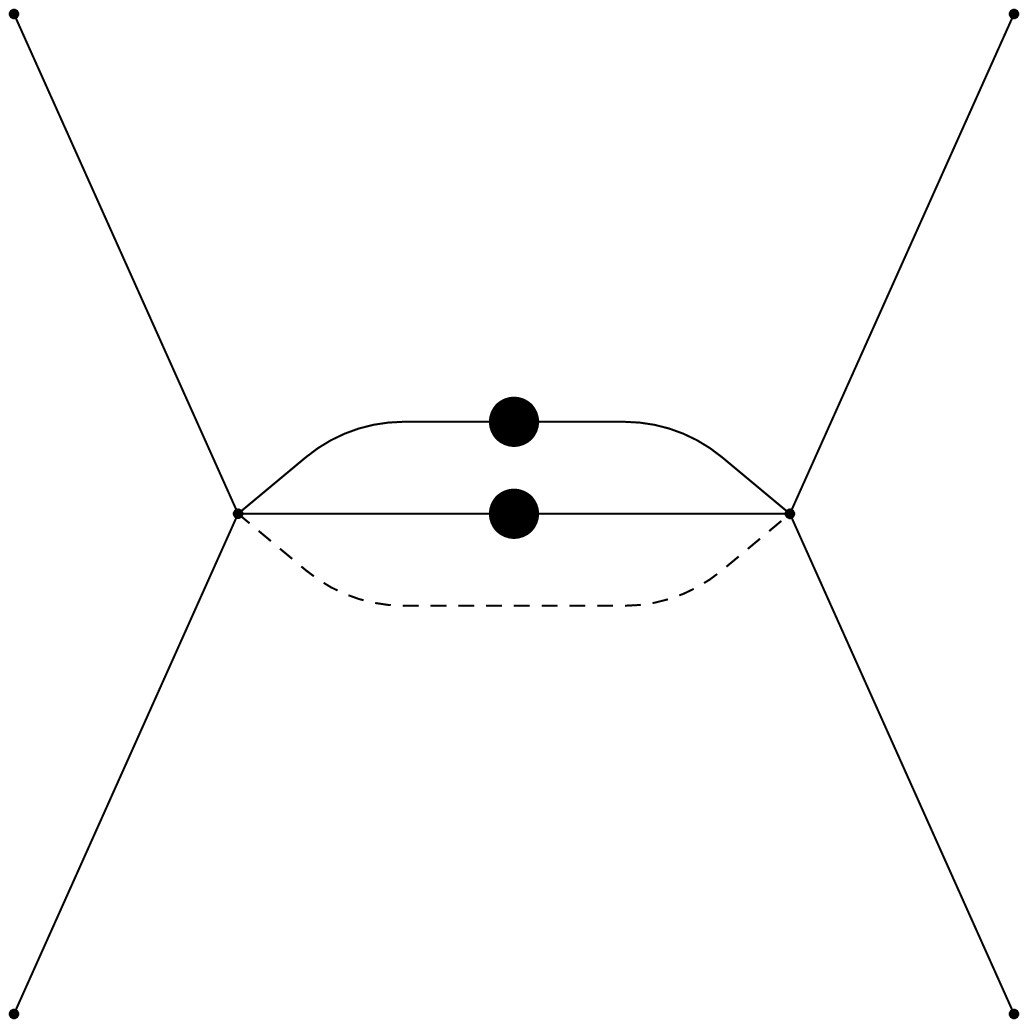}}
\subfloat[(7)]{\includegraphics[width=0.2\textwidth]{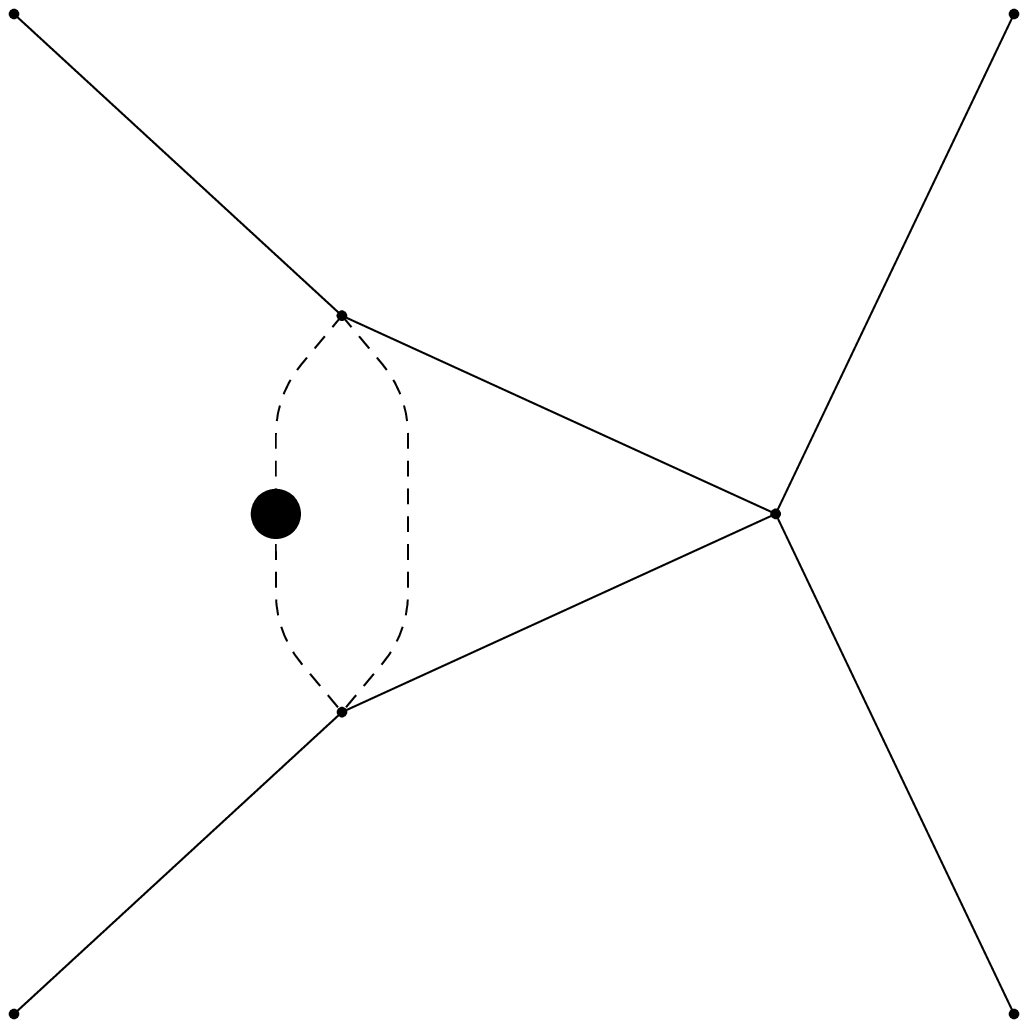}}
\subfloat[(8)]{\includegraphics[width=0.2\textwidth]{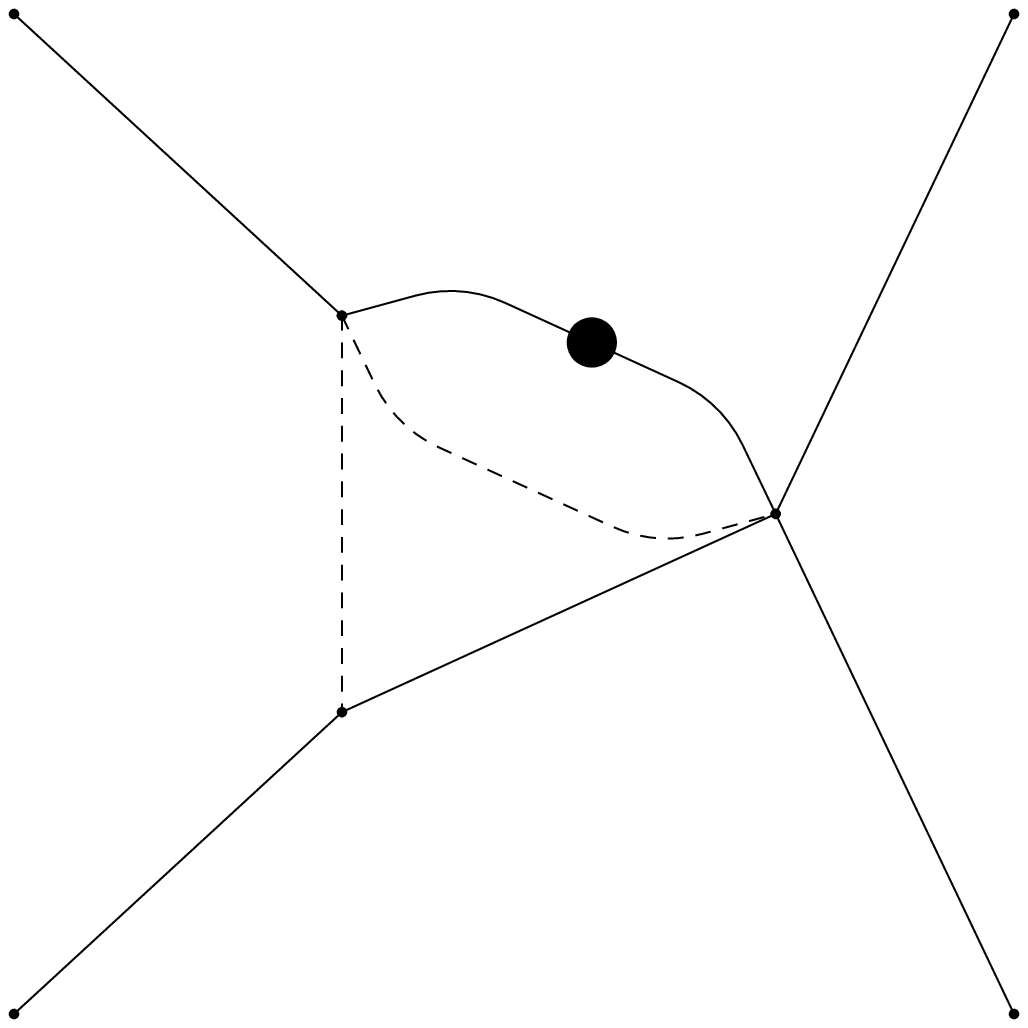}}
\subfloat[(9)]{\includegraphics[width=0.2\textwidth]{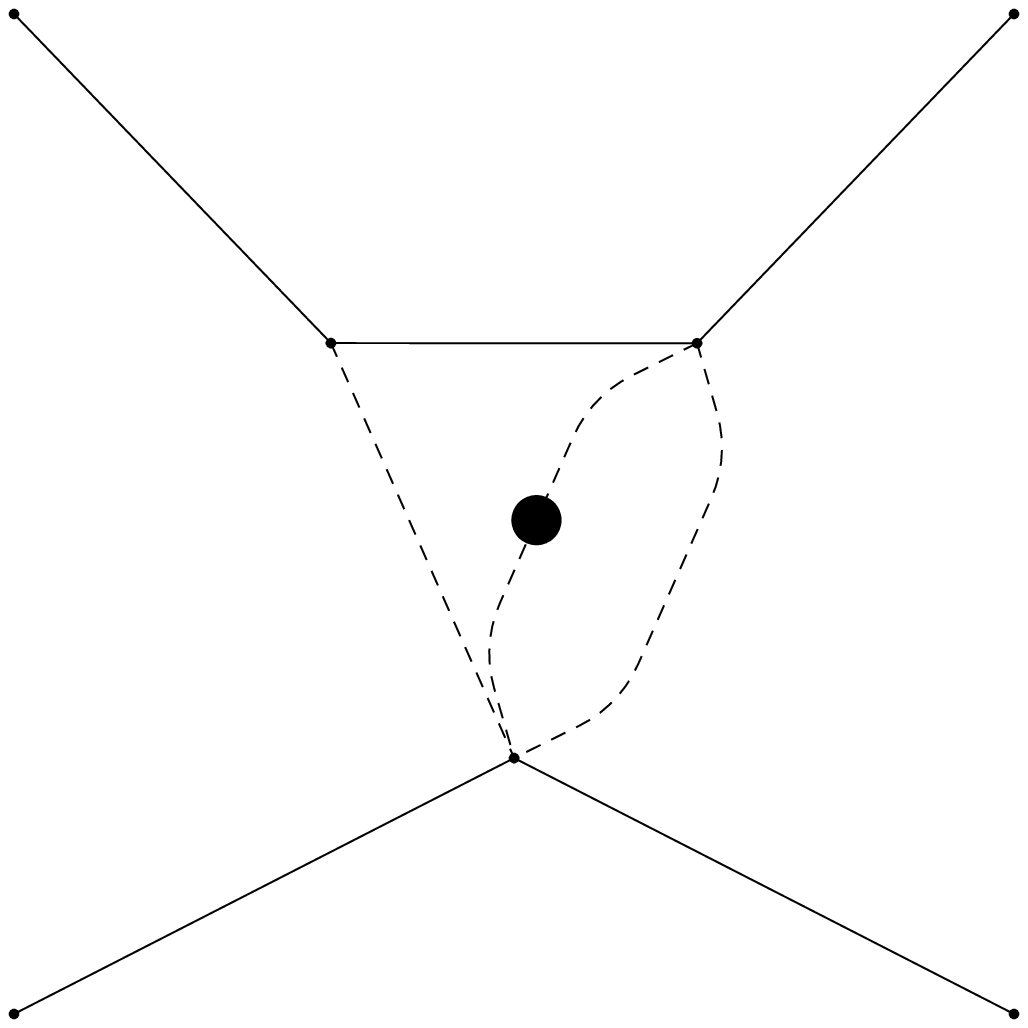}}
\subfloat[(10)]{\includegraphics[width=0.2\textwidth]{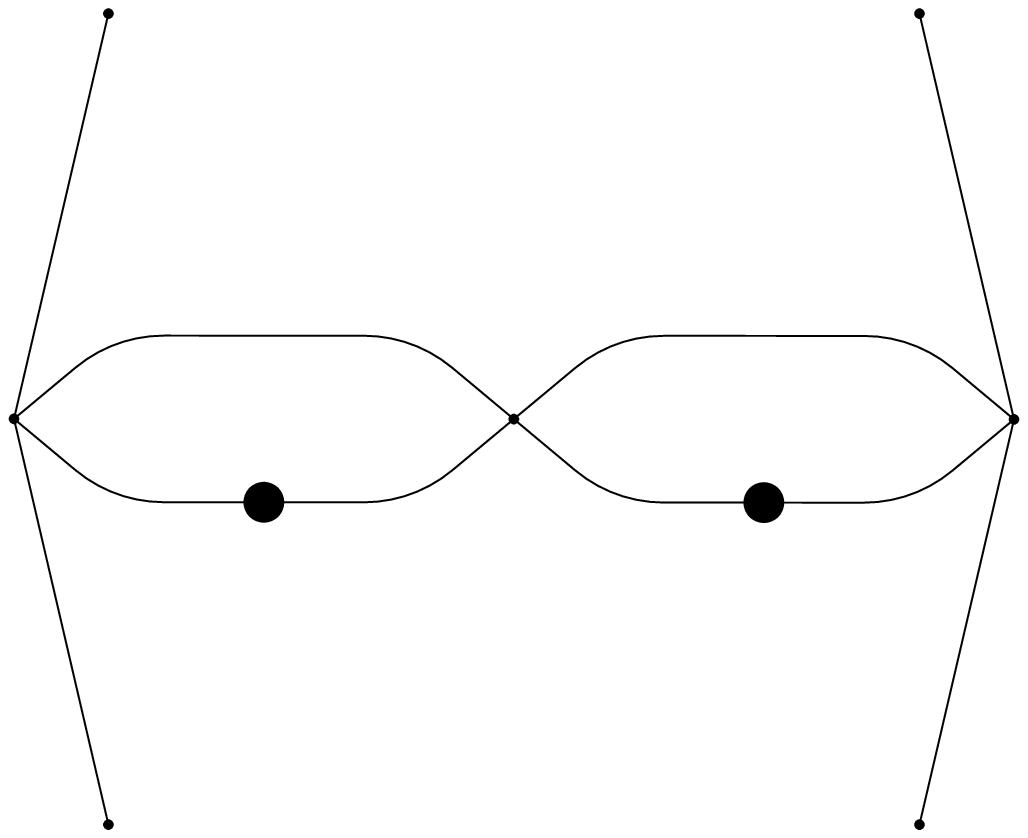}}
\newline
\subfloat[(11)]{\includegraphics[width=0.2\textwidth]{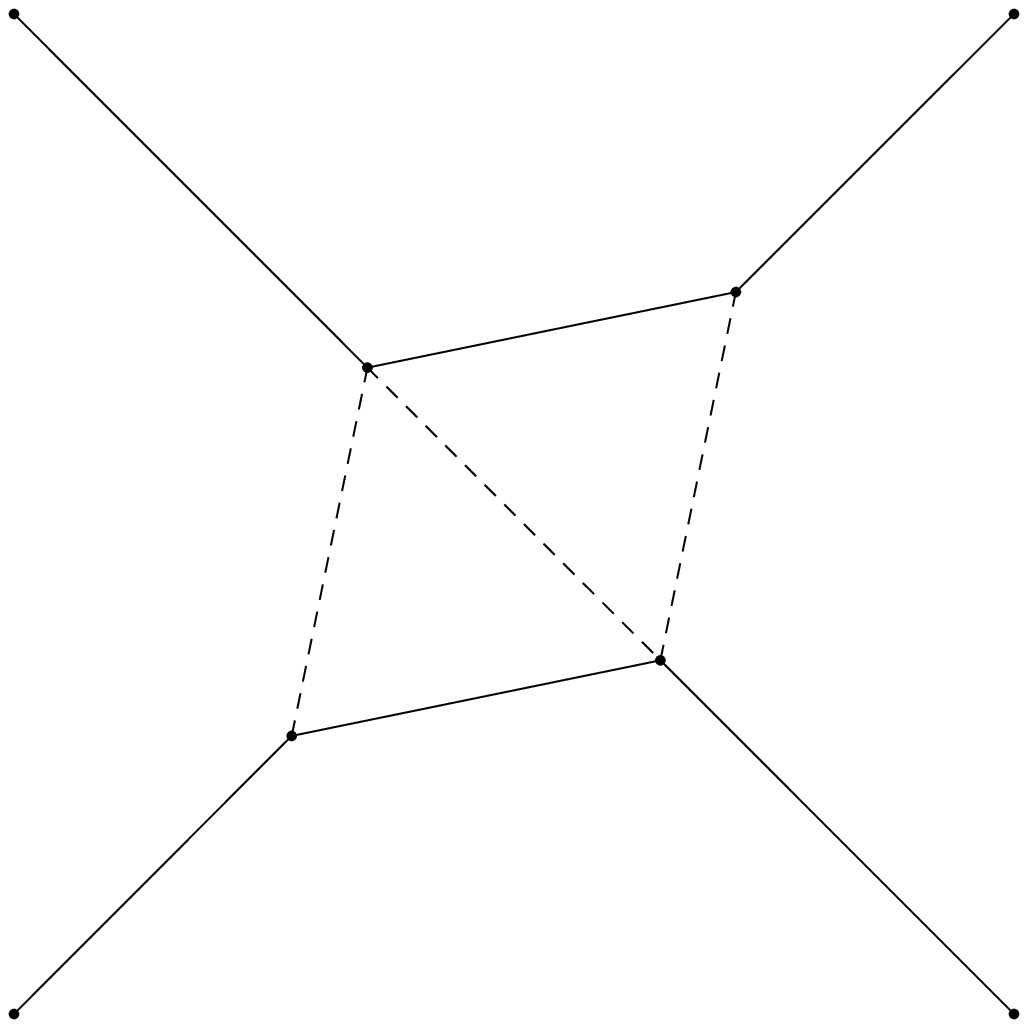}}
\subfloat[(12)]{\includegraphics[width=0.2\textwidth]{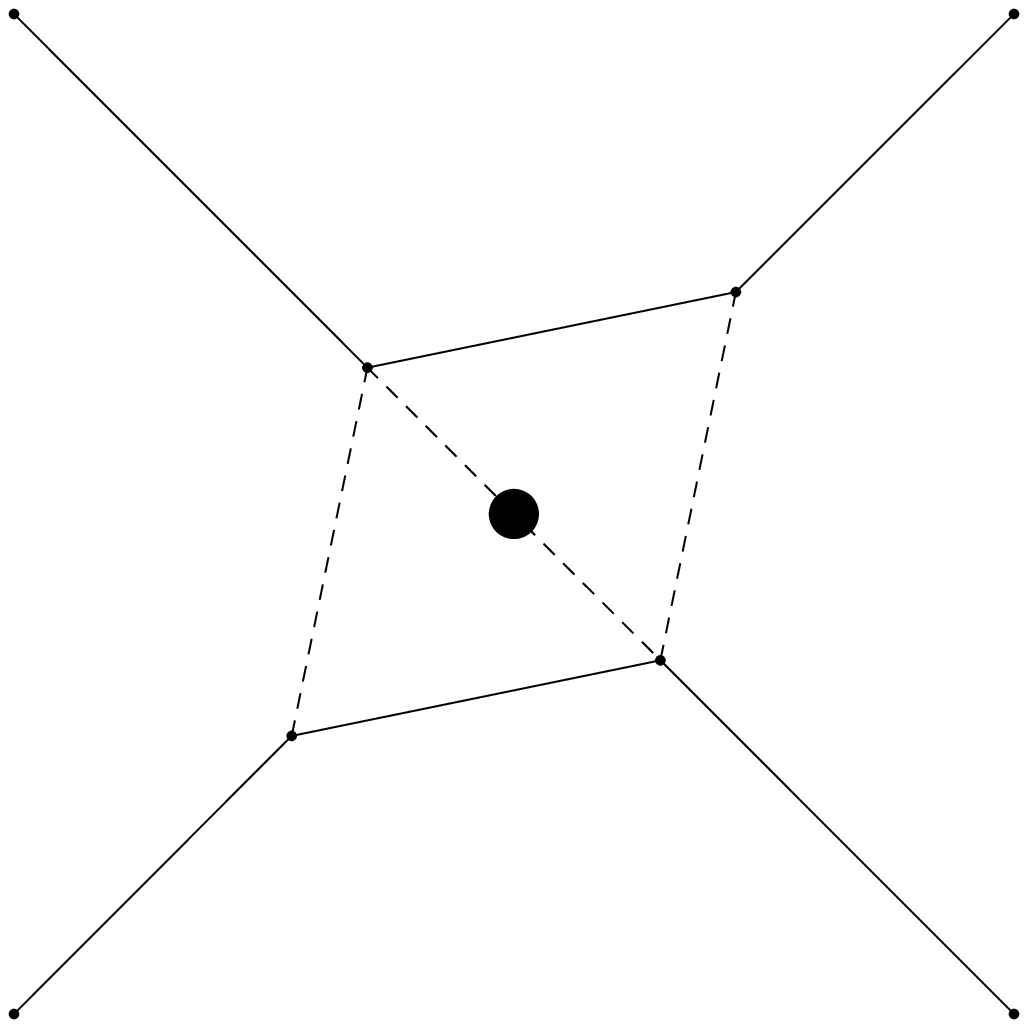}}
\subfloat[(13)]{\includegraphics[width=0.2\textwidth]{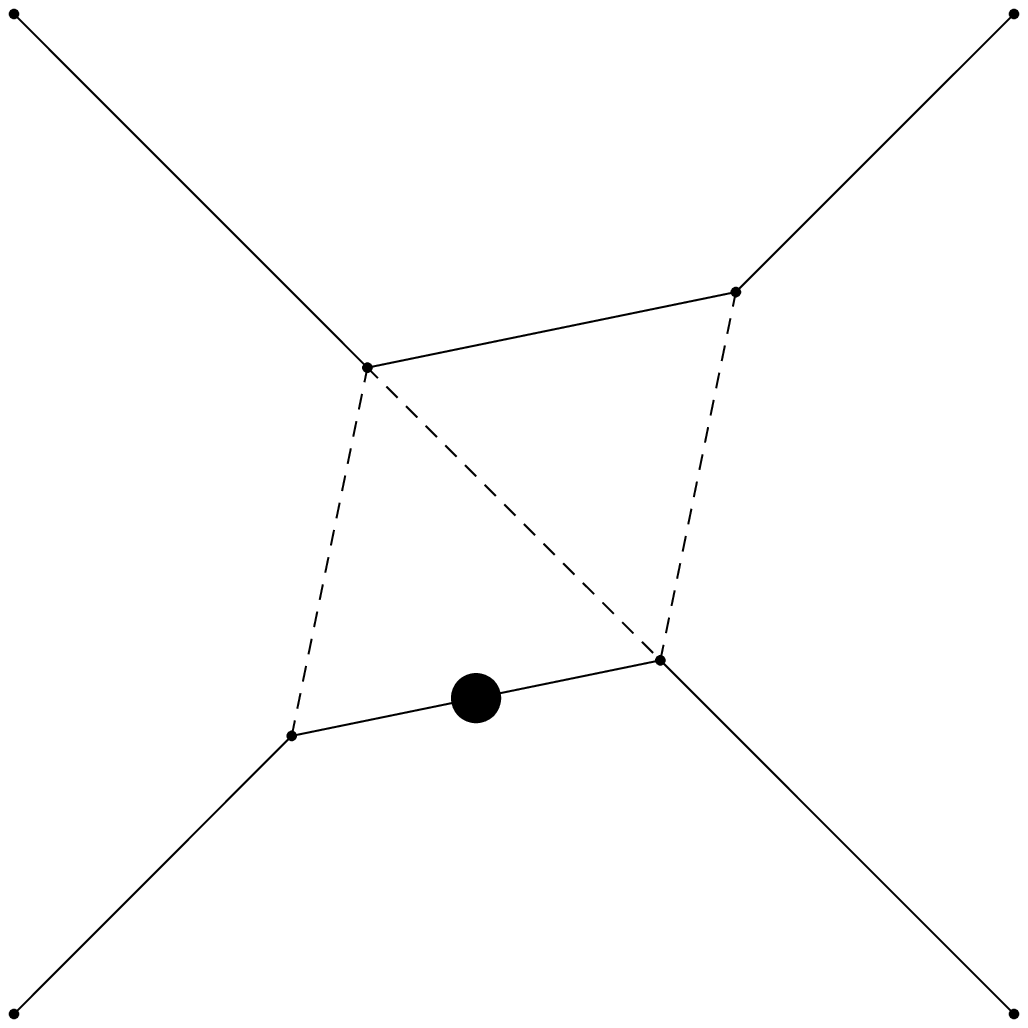}}
\subfloat[(14)${}^{\dagger}$]{\includegraphics[width=0.2\textwidth]{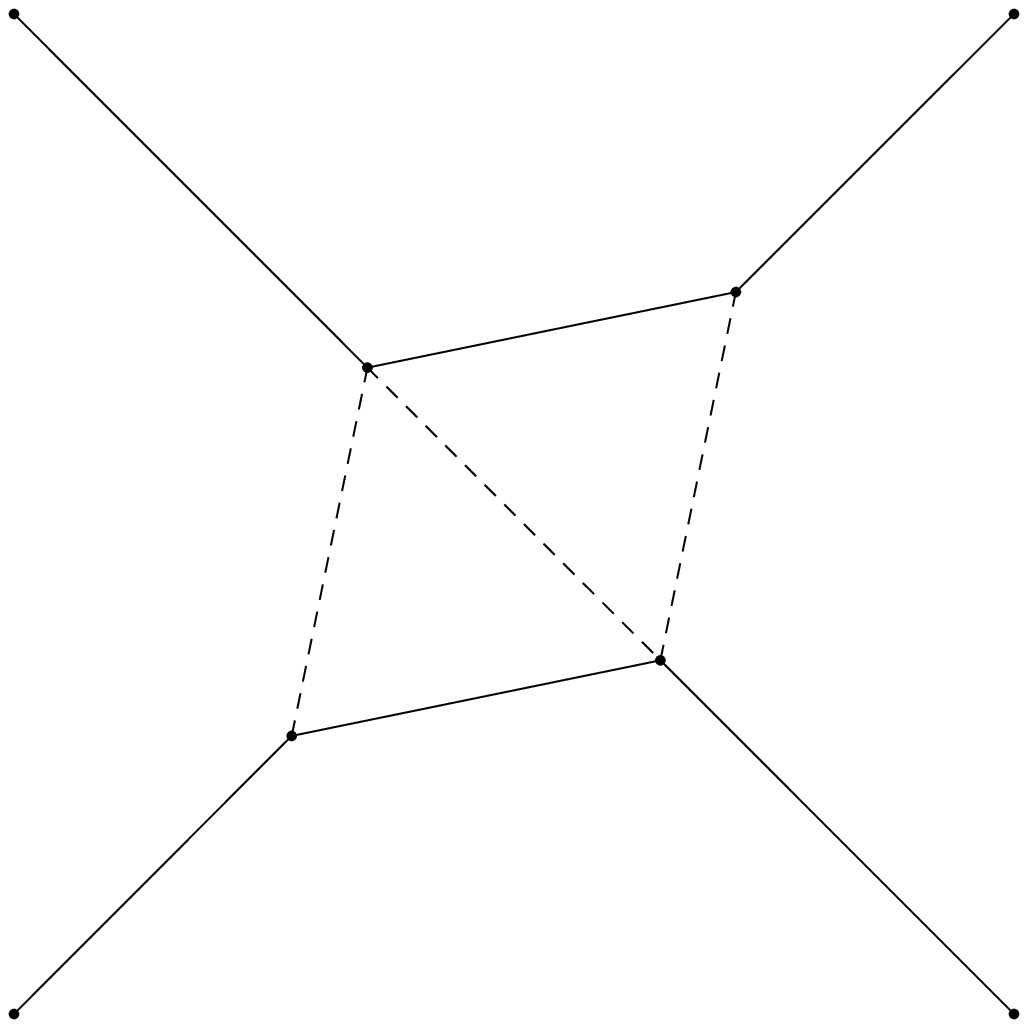}}
\subfloat[(15), (16)${}^{\dagger}$]{\includegraphics[width=0.2\textwidth]{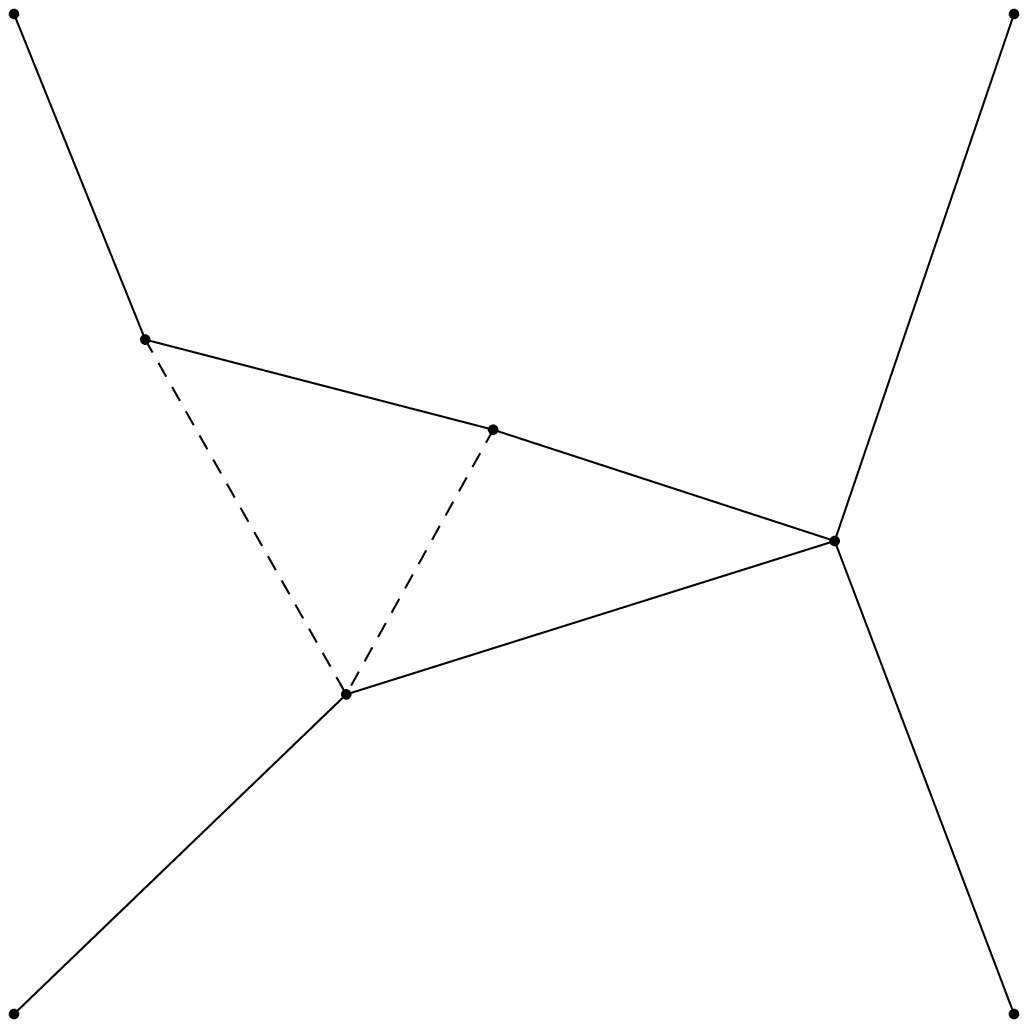}}
\newline
\subfloat[(17),(18)${}^{\dagger}$]{\includegraphics[width=0.2\textwidth]{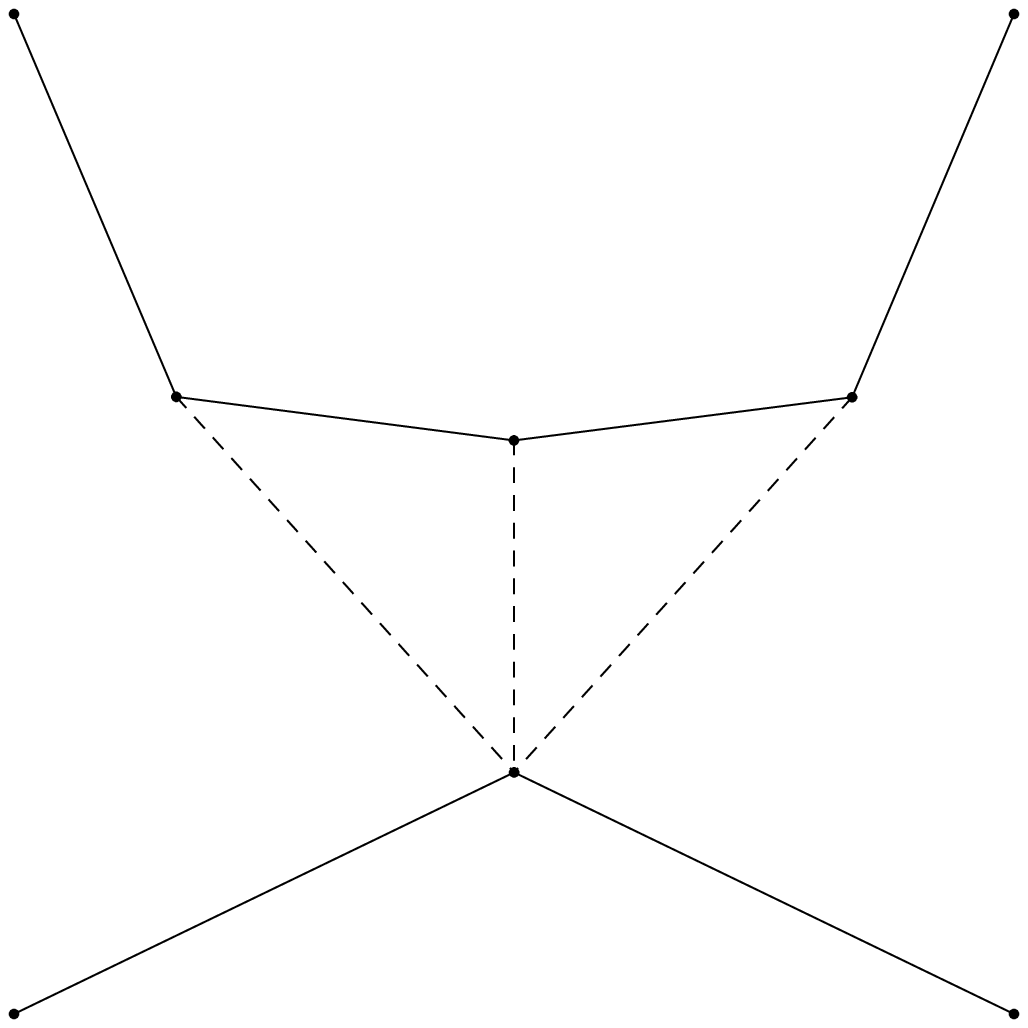}}
\subfloat[(19)]{\includegraphics[width=0.2\textwidth]{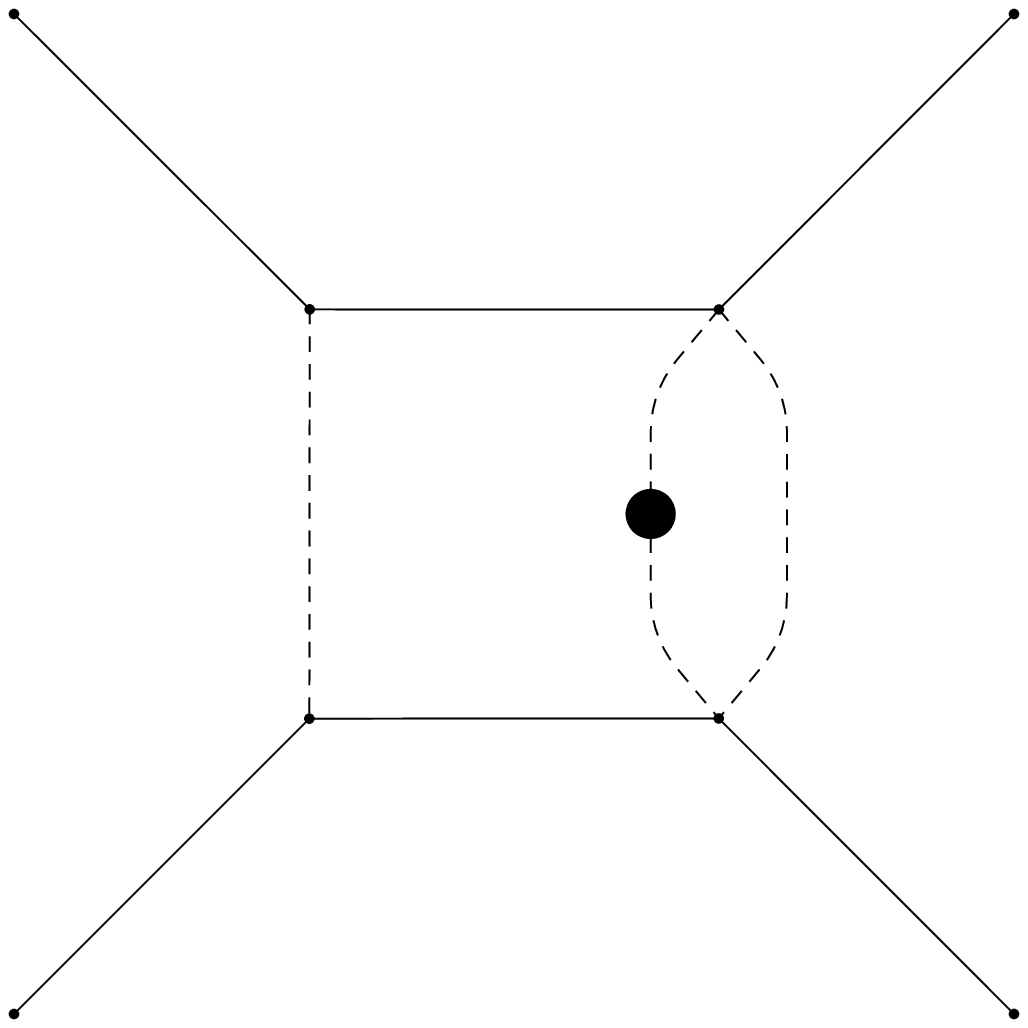}}
\subfloat[(20),(21)${}^{\dagger}$]{\includegraphics[width=0.2\textwidth]{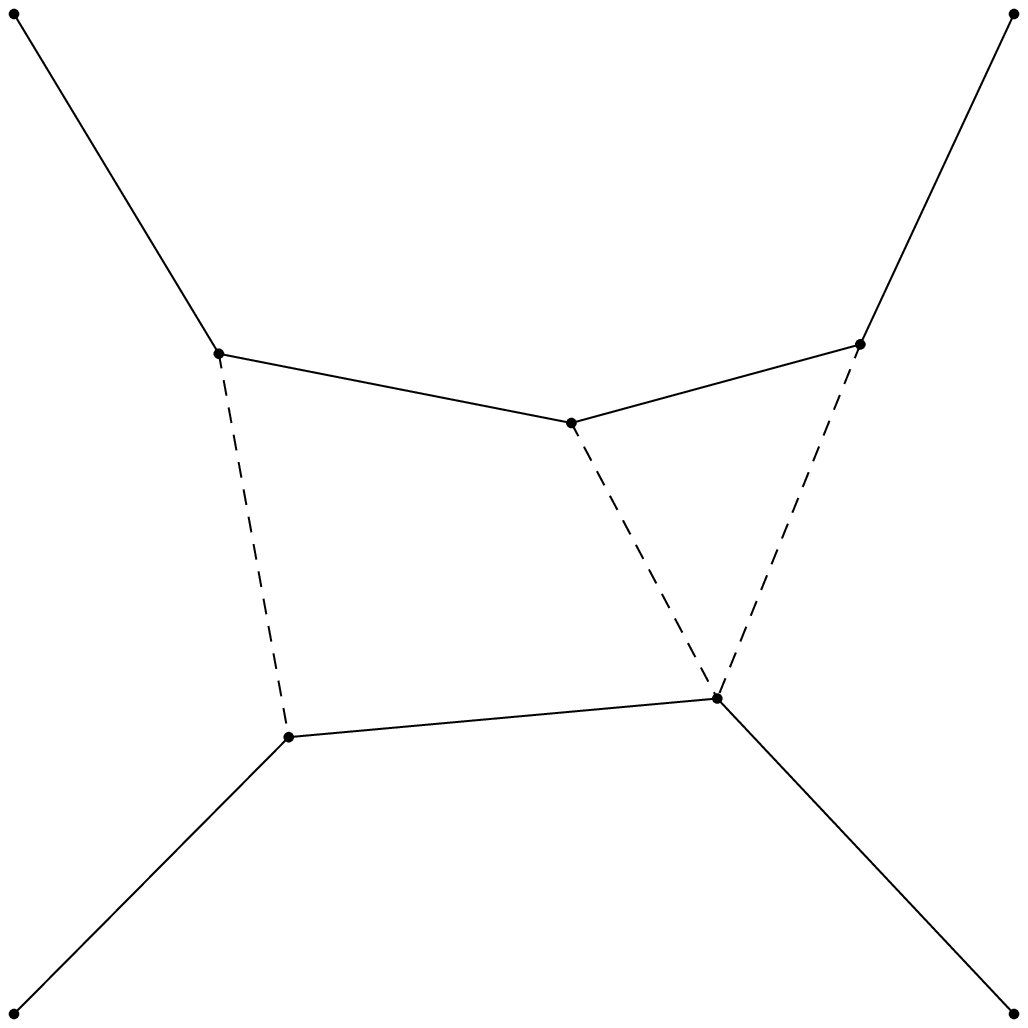}}
\subfloat[(22),(23)*]{\includegraphics[width=0.2\textwidth]{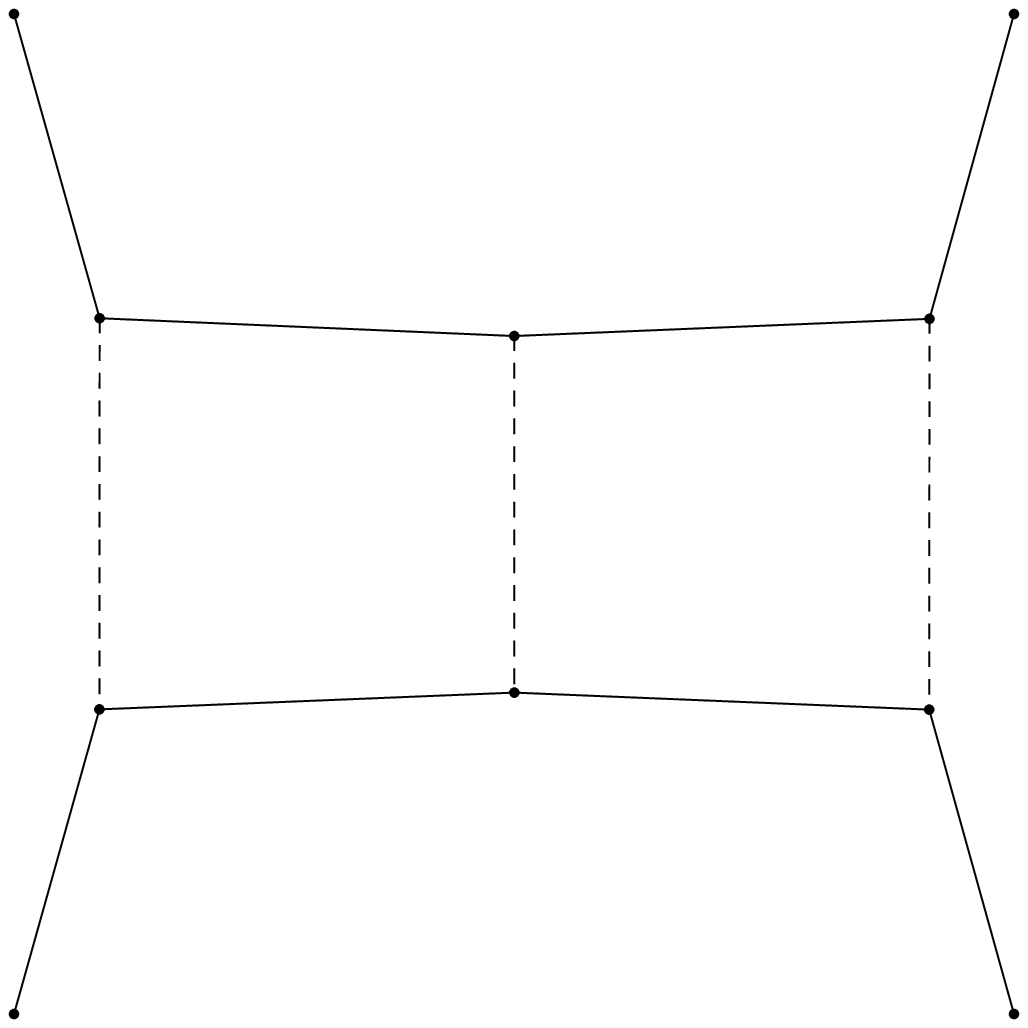}}
\caption{Master integrals for integral family (2a) of Bhabha integrals. 
Dots denote doubled propagators. 
An asterisk indicates numerators not shown in the picture, and a dagger indicates 
that the master integral is a linear combinations of integrals (possibly with dots), that are not shown.
}
\label{fig:basisBhabha2a1}
\end{center}
\end{figure}

 Let us start by presenting our basis choice for the family of integrals corresponding to Fig.~\ref{fig:bhabha2def}(2a). 
 Explicitly, we have $f_i = (m^2)^{2 \eps} \, e^{\eps \gamma_{\rm E}} \, g_{i}$, where
 \begin{align}
 g_1 =& \eps^2 \, G_{2,0,2,0,0,0,0,0,0} \,, \label{choicetwoloop1}\\
 g_2 =&\eps^2 t \, G_{0,0,0,0,2,2,1,0,0} \,, \\
 g_3 =& \frac{\eps^2 (4 \eps+1) }{\eps+1}  \, {m^2} G_{0,0,1,0,2,2,0,0,0} \,, \\
 g_4 =&\eps^2 \sqrt{-s}  \sqrt{4 {m^2}-s} \, G_{1,2,2,0,0,0,0,0,0}\,, \\
 g_5 =&\frac{1}{2} \eps^2 \sqrt{-s}  \sqrt{4 m^2 -s} (2
   G_{0,1,2,0,0,2,0,0,0}+G_{0,2,2,0,0,1,0,0,0}) \,,
   \\
   g_6 =& \eps^2 s \, G_{0,2,2,0,0,1,0,0,0} \,, \\
   g_7 =&  -2 \eps^3 \sqrt{-s}   \sqrt{4 {m^2}-s} \,G_{0,0,1,1,1,2,0,0,0}
 \,, \\
  g_8 =& -2 \eps^3 \sqrt{-s}   \sqrt{4 {m^2}-s} \, G_{0,1,2,0,1,1,0,0,0}
 \,, \\
  g_9 =& -2 \eps^3 \sqrt{-t} \sqrt{4 {m^2}-t}\, G_{1,0,0,0,1,1,2,0,0}
    \,, \\
  g_{10} =&  \eps^2 s (4 {m^2}-s) \, G_{1,2,1,2,0,0,0,0,0}  \,, \\
  g_{11} =& 4 \eps^4 \sqrt{-s-t}
   \sqrt{4 {m^2}-s-t} \, G_{0,1,1,0,1,1,1,0,0} \,, \\
  g_{12} =& -2 \eps^3 \sqrt{-s} \sqrt{4 {m^2}-s} t G_{0,1,1,0,1,2,1,0,0}
    \, \\
  g_{13} =& -2 \eps^3 {m^2} \sqrt{-t}  \sqrt{4 {m^2}-t}\,
   G_{0,2,1,0,1,1,1,0,0} \,, \\
  g_{14} =& -\frac{1}{2} \eps^2 t (\eps
   G_{0,1,1,0,1,2,1,0,0} (2 {m^2}-s)-2 {m^2}
   ({m^2} G_{0,2,2,0,1,1,1,0,0} -2 \eps
   G_{0,2,1,0,1,1,1,0,0})) \,,\\
    g_{15} =&  4 \eps^4 \sqrt{-s}   \sqrt{4 {m^2}-s} \, G_{1,0,1,1,1,1,0,0,0}
 \,, \\
  g_{16} =& -\frac{1}{4} \eps^2 (4 {m^2}-s) \Big(-8
   \eps^2 G_{1,0,1,1,1,1,0,0,0}+8 \eps
   {m^2} G_{1,0,1,2,1,1,0,0,0} \nonumber \\ &
   +8 \eps
   {m^2} G_{1,0,2,1,1,1,0,0,0}+4 \eps
   G_{0,0,1,1,1,2,0,0,0}-2 \eps
   G_{0,1,2,0,1,1,0,0,0}\nonumber \\&+2
   G_{0,1,2,0,0,2,0,0,0}+G_{0,2,2,0,0,1,0,0,0}\Big) \,, \\
  g_{17} =& 4 \eps^4 \sqrt{-t}   \sqrt{4 {m^2}-t} \, G_{1,0,1,0,1,1,1,0,0}
  \,, \\
  g_{18} =&  \eps \Big[  t \left(2 \eps^3
   G_{1,0,1,0,1,1,1,0,0}+(2 \eps+1)
   {m^4} G_{1,0,2,0,2,1,1,0,0} \right) \nonumber \\& +2
   \eps^2 G_{1,0,0,0,1,1,2,0,0} (3
   {m^2}-t)+\eps {m^2}
   G_{0,0,1,0,2,2,0,0,0} \Big]\,, \\
  g_{19} =&-2 \eps^3 \sqrt{-s}   \sqrt{4 {m^2}-s} t\, G_{1,1,0,0,1,1,2,0,0}
  \,, \\
  g_{20} =&  4 \eps^4 \sqrt{-s} \sqrt{-t} \sqrt{4 {m^2}-s}
   \sqrt{4 {m^2}-t} \,
   G_{1,1,1,0,1,1,1,0,0}\,, \\
  g_{21} =&\eps^2  \sqrt{-s}  \sqrt{4 {m^2}-s} \,  \Big(-8
   \eps^2  t \, G_{1,1,1,0,1,1,1,0,0}
 +8
   \eps^2 \, G_{0,1,1,0,1,1,1,0,0} \nonumber \\ &
    -4
   \eps (2 \eps-1)  \, G_{1,1,1,0,1,1,1,-1,0} 
    \Big)  \,, \\ 
g_{22} =& 4 \eps^4 s t (4 {m^2}-s) \, G_{1,1,1,1,1,1,1,0,0}  \,, \\
g_{23} =& 4 \eps^4 s (4{m^2}-s)\, G_{1,1,1,1,1,1,1,-1,0}  \label{choicetwoloop23}  \, .   
 \end{align}
The corresponding Feynman diagrams can be found in Fig.~\ref{fig:basisBhabha2a1}. 

%
%
%
%
%

\subsection{System of differential equations}

We find that the basis choice of eqs. (\ref{choicetwoloop1})--(\ref{choicetwoloop23})  puts the differential equations into canonical form, i.e.
\begin{align}\label{diffeq2loop}
d\, f = \eps\, d \tilde{A} \, f \,.
\end{align}
Here $f$ contains $23$ integrals, and $\tilde{A}$ is a $23\times 23$ matrix.
The matrix $\tilde{A}$ can be put into the following form,
\begin{align}\label{matrixAtilde2loop}
\tilde{A} =& B_1 \, \log(x) + B_2 \, \log(1+x) + B_3 \, \log(1-x) + B_4 \, \log(y) + B_{5} \, \log(1+y) \,\nonumber \\
& + B_6 \log(1-y)\ + B_{7} \log(x+y) + B_{8} \log(1+x y) \,\nonumber \\
& + B_9 \log(x + y - 4 x y + x^2 y + x y^2) + B_{10} \log\left( \frac{1+Q}{1-Q} \right)  \nonumber \\
&+B_{11} \log\left(  \frac{(1+x) + (1-x) Q}{(1+x)-(1-x)Q} \right) +B_{12} \log\left( \frac{(1+y) + (1-y) Q}{(1+y)-(1-y)Q} \right) \,,
\end{align}
Here 
\begin{align}
Q =\sqrt{ \frac{ (x+y)(1+ x y)}{x+y-4 x y+x^2 y +x y^2}} \,,
\end{align}
and the $B_{i}$ are constant matrices, see Appendix \ref{appendix_bi}.

It is important to note that the matrix $\tilde{A}$ only contains logarithmic singularities, so that
we can expect the answer to be expressed in terms of iterated integrals.

We see that w.r.t. the one-loop case, the four terms in the last two lines of eq. (\ref{matrixAtilde2loop}) are new.
 Analyzing the arguments of the logarithms we find that most of them only have singularities at $x=0$, $y=0$, $x=1$, $y=1$.
However, the coefficient of $B_{9}$ has additional singularities that take a more complicated form in terms of $x$ and $y$. In
they original variables, they correspond to $s=-t$. 
It is worth pointing out that the matrix $B_{9}$ only has one non-zero entry, see eq. (\ref{matrixb9}),
and this symbol entry will appear first at degree five for integral $11$, and at higher degrees for other integrals.

%
%
%
%
%

\subsection{Solution at symbol level}
\label{sec:symbol2}

We find that we can determine all integration constants from physical conditions, up to a few
trivial propagator-type integrals.
In particular, we find for the first order in the $\eps$ expansion
\begin{align}
f^{(0)} = \{1, -1, -1/2, 0, 0, 0, 0, 0, 0, 0, 0, 0, 0, -1/
  2, 0, 0, 0, -1, 0, 0, 0, 0, 0 \} \,.
\end{align}
This, together with the differential equation (\ref{diffeq2loop}) and (\ref{matrixAtilde2loop}), completely specifies the symbol.

Before writing down the explicit solution for the basis integrals,
we find it useful to analyze the symbols of the functions.
The reason is that although one would expect that 
all terms appearing in eq. (\ref{matrixAtilde2loop})
eventually enter the r.h.s. of the differential equations for the integrals with many
propagators, the matrices $B_{i}$ are sparse matrices with many zero entries.
It is therefore possible that, at least to some low order in the $\eps$ expansion,
not all of the structures present in general will be needed.

Having physical applications at NNLO in mind, we explicitly analyze this question
up to degree four. We can make the following structural observations:
\begin{itemize}
\item At order $\eps, \eps^2$ and $\eps^3$, the symbol alphabet remains the same as at one loop.
\item At order $\eps^4$ all functions except $f_{11}$ have the same symbol alphabet as at one loop.
\end{itemize}
In other words, most of the new symbol entries, corresponding to matrices $B_{9}, B_{10}, B_{11}, B_{12}$ given in eqs.  (\ref{matrixb9})- (\ref{matrixb12}), are not needed for physical two-loop results!
(It would be interesting to know whether in our basis, the weight four part of integral $11$ is needed in physical calculations.)

%
%
%
%
%

\subsection{Explicit results in terms of Goncharov polylogarithms}

Having observed that the symbol alphabet for all functions except for $f_{11}$ is the same as at one loop,
all the observations made there apply immediately to this case. In particular, as explained in section \ref{solGoncharov},
we can write the analytic answer in terms of certain Goncharov polylogarithms. 
One may call these functions two-dimensional harmonic polylogarithms.

All analytic results are collected in the ancillary file {\tt Results_2a.m} which is supplied with the paper.
Here, for illustration, we present the result for integral $f_{23}$, up to weight four.
It reads
\begin{align}
f_{23} =& \eps^2 \Big[  -12 H_{0,0}(x) \Big] + \eps^3 \Big[  -16 G_0(y) H_{0,0}(x)+32 G_1(y) H_{0,0}(x)+8 H_{2,0}(x) 
\nonumber \\ & +16 H_{-1,0,0}(x)-4 H_{0,0,0}(x)+\frac{4}{3} \pi ^2 H_0(x)+4
   \zeta_3 \Big] + \eps^4 \Big[
   32 G_0(y) H_{-2,0}(x)
   \nonumber \\ & 
   -32 H_{-2,0}(x) G_{-\frac{1}{x}}(y)-32 H_{-2,0}(x) G_{-x}(y)+64 G_{1,0}(y) H_{0,0}(x)-128
   G_{1,1}(y) H_{0,0}(x)
   \nonumber \\ & 
   -32 H_{0,0}(x) G_{-\frac{1}{x},0}(y)+64 H_{0,0}(x) G_{-\frac{1}{x},1}(y)-32 H_{0,0}(x)
   G_{-x,0}(y)\nonumber \\ & 
   +64 H_{0,0}(x) G_{-x,1}(y)-16 H_0(x) G_{-\frac{1}{x},0,0}(y)+32 H_0(x) G_{-\frac{1}{x},0,1}(y)
   \nonumber \\ & +16
   H_0(x) G_{-x,0,0}(y)-32 H_0(x) G_{-x,0,1}(y)+64 G_0(y) H_{-1,0,0}(x)
   \nonumber \\ & -64 H_{-1,0,0}(x) G_{-\frac{1}{x}}(y)-64
   H_{-1,0,0}(x) G_{-x}(y)\nonumber \\ & 
   -48 G_0(y) H_{0,0,0}(x)+48 H_{0,0,0}(x) G_{-\frac{1}{x}}(y)+48 H_{0,0,0}(x) G_{-x}(y)-120
   H_{-3,0}(x)\nonumber \\ & 
   +\frac{52}{3} \pi ^2 H_{0,0}(x)+48 H_{3,0}(x)+128 H_{-2,-1,0}(x)-120 H_{-2,0,0}(x)-48 H_{-2,1,0}(x)
   \nonumber \\ & +64
   H_{-1,-2,0}(x)-32 H_{-1,2,0}(x)-48 H_{2,-1,0}(x)+32 H_{2,0,0}(x)+16 H_{2,1,0}(x)\nonumber \\ & 
   +64 H_{-1,-1,0,0}(x)-80
   H_{-1,0,0,0}(x)+76 H_{0,0,0,0}(x)+\frac{8}{3} \pi ^2 G_0(y) H_0(x)\nonumber \\ & 
   -\frac{40}{3} \pi ^2 H_0(x)
   G_{-\frac{1}{x}}(y)+8 \pi ^2 H_0(x) G_{-x}(y)-16 \zeta_3  H_{-1}(x)-28 \zeta_3 H_0(x)\nonumber \\ & +\frac{8}{3} \pi ^2
   H_{-2}(x)-\frac{4}{3} \pi ^2 H_2(x)-\frac{4 \pi ^4}{15}
    \Big] + \cO(\eps^5)
\end{align}
This formula may be evaluated numerically using standard tools available in the literature, see e.g. \cite{Bauer:2000cp}.

We checked our results against various formulas available in the literature, e.g. \cite{Czakon:2004wm}.
A particularly non-trivial test the agreement with the result of \cite{Smirnov:2001cm} for integral $f_{22}$,
since the differential equations for this integral involve most other integrals.
We also performed numerical checks in the Euclidean region by the implementation of sector decomposition \cite{Binoth:2003ak,Borowka:2012yc}
in FIESTA \cite{Smirnov:2008py,Smirnov:2009pb}, 
and analytic checks in asymptotic limits that were obtained from Mellin-Barnes representations.

%
%
%
%
%

\subsection{Explicit parametrization for Chen iterated integral}
\label{int11}

In the previous paragraph, we have written down an explicit solution for all integrals
except for $f_{11}$ up to degree four. We also explained, in section \ref{sec:chen}, how to
write the general solution to the two-loop problem in terms of Chen iterated integrals.

Let us exemplify this using $f_{11}$. The basis integral is finite as $\eps \to 0$, 
which explains why it first appears at order $\eps^4$ in our normalization.

Using the explicit results for all other integrals, and expanding eq. (\ref{diffeq2loop}) to fourth order in $\eps$,
we find that $f^{(4)}_{11}$ satisfies the following differential equation,
\begin{align}\label{diff11}
d\, f^{(4)}_{11} = g_1 \, d\, \log \left(\frac{1-Q}{1+Q}\right) + g_2 \, d\, \log \left(
 \frac{(1+x) + (1-x) Q}{(1+x)-(1-x)Q}
\right) + g_3 \, d\, \log \left(
 \frac{(1+y) + (1-y) Q}{(1+y)-(1-y)Q} 
\right) \,,
\end{align}
where the three functions $g_1, g_2, g_3$ are given explicitly by
 \begin{align}
 g_1 =& -16 G_0(y) H_{0,0}(x)+32 G_1(y) H_{0,0}(x)+8 G_{0,0,0}(y)-16 G_{0,0,1}(y)-32 H_{-2,0}(x)\nonumber \\
 &+8 H_{0,0,0}(x)  -16
   H_{1,0,0}(x)+\frac{8}{3} \pi ^2 G_0(y)+\frac{16}{3} \pi ^2 G_1(y)-\frac{8}{3} \pi ^2 H_0(x)-32 \zeta_3 \,, \\
   g_2 =&-16 H_{0,0,0}(x)-\frac{8}{3} \pi ^2 H_0(x)\,, \\
   g_3 = & \, 8 H_0(x) G_{-\frac{1}{x},0}(y)-16 H_0(x) G_{-\frac{1}{x},1}(y)-8 H_0(x) G_{-x,0}(y)+16 H_0(x) G_{-x,1}(y) \nonumber \\ & +16 H_{-1,0}(x)
   G_{-\frac{1}{x}}(y)-16 H_{-1,0}(x) G_{-x}(y)-8 G_0(y) H_{0,0}(x)-8 H_{0,0}(x) G_{-\frac{1}{x}}(y)\nonumber \\ &+8 H_{0,0}(x) G_{-x}(y)+8
   G_{-\frac{1}{x},0,0}(y)-16 G_{-\frac{1}{x},0,1}(y)+8 G_{-x,0,0}(y)-16 G_{-x,0,1}(y)\nonumber \\ &-16 G_{-1,0,0}(y)+32 G_{-1,0,1}(y)-16
   H_{-2,0}(x)+8 H_{0,0,0}(x)+\frac{20}{3} \pi ^2 G_{-\frac{1}{x}}(y)\nonumber \\ &+4 \pi ^2 G_{-x}(y)-\frac{32}{3} \pi ^2
   G_{-1}(y)-\frac{4}{3} \pi ^2 H_0(x)-24 \zeta_3 \,.
 \end{align}
They can also be written in terms of classical polylogarithms, 
in a form that is manifestly real-valued in the region $0<x<1,0<y<1$.

At $x=1$, equation (\ref{diff11}) simplifies, and integrating back we find
\begin{align}
f^{(4)}_{11}(x=1,y) = \frac{8}{3} \pi ^2  H_{0,0}(y)+8  H_{0,0,0,0}(y)-\frac{16}{3} \pi ^2 H_2 (y)+16 H_4 (y)+\frac{32 \pi ^4}{45}\,.
\end{align}
Note that in particular, at $y=1$, we have $f^{(4)}_{11}(x=1,y=1)=0$.
We may also note that $f^{(4)}$ vanishes for $x=-y$.

As we reviewed in section \ref{sec:chen}, we can always write the solution to an equation 
like (\ref{diff11}) in terms of Chen iterated integrals \cite{Chen1997}.\footnote{Strictly speaking, the answer below
is given in a hybrid formulation, since we are using the 2dHPL form of the functions $g_1 , g_2 , g_3 $. We could
of course represent the latter also as integrals over logarithmic differential forms.}
In order to do this, we choose a base point (corresponding to a boundary condition) in the kinematical space.

We may choose different base points depending on the application we have in mind.
For example, for numerical evaluation in the Euclidean region $0<x<1, 0<y<1$,
choosing $(x,y)=(1,1)$ as a base point is natural, and we have $f^{(4)}_{11}(x=1,y=1)=0$.

To evaluate  $f^{(4)}_{11} (x_f, y_f)$ at some point $(x_f, y_f)$, we integrate the differential form of eq. (\ref{diff11})
along a contour. The contour is chosen such that it avoids singularities and branch cuts. For values $0<x_f<1, 0<y_f<1$, a possible choice for the contour $\mathcal{C}$ is, cf. Fig.~\ref{fig:paths},
\begin{align}
x(s) = (1-s) + s x_f \,, \qquad y(s) = (1-s) + s x_f \,.
\end{align}
Then, we have
\begin{align}
f^{(4)}_{11}(x_f, y_f) =  \int_{\mathcal{C}} \, d \, f^{(4)}_{11} \,.
\end{align}
Of  course, one can also choose more complicated contours.

In this way, we can easily obtain numerical values in the region $0<x<1,0<y<1$. 
(For other region, one has to define a contour that avoids the branch cuts.)
For example, we find
\begin{align}
f^{(4)}_{11}(1/3, 2/5 ) =& 108.10505928259012 \,,\\
f^{(4)}_{11}(6/7, 3/5 ) =& 46.46787346208666 \,,\\
f^{(4)}_{11}(3/8, 7/9 ) =& 130.0624797134173 \,.
\end{align}

For numerical evaluation in the physical region $x>0, y<0$ (or $x>0, y<0$), it is convenient to choose a different 
base point. $x=-y$, e.g. $x=1,y=-1$ is a good choice. 
In this way, we straightforwardly obtain numerical values in that region.

A similar type of integral representation was used in ref. \cite{Dixon:2011nj} for the evaluation
of two-loop six-point amplitudes in $\cN=4$ SYM.

It would be interesting to know whether the weight four function $f^{(4)}_{11}$ can be rewritten in
terms of classical polylogarithms and ${\rm Li}_{2,2}(x,y)$, see the discussion of section \ref{solmultiple}.
Unfortunately, even if such a representation does exist, it
is by no means guaranteed that the individual terms in such an expression have a symbol alphabet
given by eq. (\ref{matrixAtilde2loop}). Instead, it could be that `spurious' symbol entries are needed
for such a representation, with the latter canceling in the complete answer. In that case, 
the main difficulty lies in finding these spurious symbol entries in a systematic way.

%
%
%
%
%

\section{Discussion and outlook}
\label{sec:outlook}

In this paper, we analytically computed the master integrals for all Feynman integrals
of the family of eq. (\ref{deftwoloop2a}), corresponding to Fig.~\ref{fig:bhabha2def}(2a).
This completes the previous partial results for some of the integrals available in
the literature.
We wrote our results explicitly up to weight four, which is sufficient for NNLO applications.
We also identified the class of functions needed to any order in $\eps$, and higher
order results can be obtained by expanding the general solution of eq. (\ref{path_ordered}),
taking into account the boundary conditions discussed in section \ref{sec:analytic}.
Moreover, as a corollary, the symbol at any order in $\eps$ can be trivially obtained, as explained in 
section \ref{sec:symbol2}.

We would like to discuss further the nature of our results for the master integrals.
Let us observe that, usually, a result for some function, e.g. a Feynman integral,
is characterized as analytical if it is explicitly expressed in terms of known functions.
By known functions we usually mean elementary functions, and sets of special
functions whose properties are well studied, and that arise in many different problems.
For a discussion of this topic, see \cite{wolfram}.
The latter criteria are obviously important, because otherwise one could
introduce a new special function for each problem, and declare the problem solved.
Clearly, this would not be satisfactory. 
What can we say about our results for the master integrals from this point of view?

We would like to argue that Chen iterated integrals can be considered as a 
class of special functions that appear in many physical problems, particularly
in Feynman integrals. 
They are well studied in the mathematical literature,
and important properties, such as their singularity structure, monodromies,
differential equations, functional relations, can be studied in a simple way.
They also appear in other areas, such as the theory of knot invariants \cite{Kontsevich}.
In the case where the logarithmic integration forms have rational arguments, one obtains
functions that arise frequently in quantum field theory computations, such as, in
decreasing order of complexity, Goncharov polylogarithms, 2dHPLs, HPLs,
polylogarithms, and logarithms.

Just like many other special functions, these functions appear as the solution of a 
class of differential equations. This is also the context in which they appear here.
Let us emphasize that these iterated integrals form exactly the class of functions needed to 
describe the given family of Feynman integrals, to all orders in $\epsilon$.
Therefore they are the natural language to discuss them.

We should mention, as discussed in section \ref{solmultiple}, that if one is interested in writing
down the solution to a given order in $\eps$, one can try to write the solution in
terms of a minimal number of functions, such as classical polylogarithms for degree two,
at the cost of having more complicated arguments of those functions.
While it is certainly desirable to write an answer in terms of the smallest possible
number of building blocks, we wish to point out that such a rewriting may make less obvious some
of the properties that are transparent in the more general language of iterated integrals. 
In particular, there is no canonical way of presenting the answer due to
numerous identities between polylogarithms of different arguments, see e.g. \cite{LewinPolylogarithms}.
This is probably the reason why in the theoretical physics literature on Feynman integrals, 
such a rewriting is often not performed, even in the cases where it is rather straightforward.

Given the physical relevance of the class of functions defined from the symbol
alphabet (\ref{symbolalphabet1}) in this and other problems, 
it would be interesting to construct a set of single-valued 
functions of this type. 
This is a topic that is very well-studied in the mathematical
literature, see  \cite{iterated1,iterated2} and references therein.
Likewise, one could study the generalization to (\ref{matrixAtilde2loop})
from the same point of view. As we emphasized, this case is the appropriate
class of functions for the solution to all orders in $\eps$.

In this context we wish to mention that there is a choice of variables that
suggests that this class of functions is natural.
As in \cite{Bonciani:2004gi} we can introduce a third variable $z$ in analogy with eq. (\ref{defxy}),
namely ${-u}/{m^2} = {(1-z)^2}/{z}$.
The variables $x,y,z$ are of course constrained due to $u=-s-t+4 m^2$, which becomes
\begin{align}\label{constraintxyz}
\frac{(1-x)^2}{x}+\frac{(1-y)^2}{y}+\frac{(1-z)^2}{z} + 4 = 0\,.
\end{align}
Using these constrained variables, the symbol alphabet takes the following simple form,
\begin{align}
\{  x \,, \; 1\pm x\,, \; y \,,  \; 1\pm y \,, \; z \,, \; 1 \pm z\,, \;   x+y \,,\;  1+ x y \,, \; x+z \,, \; 1+ x z \,, \; y + z \,, \;1+ y z \} \,.
\end{align}
Finally, we wish to mention that yet another equivalent representation is obtained by letting $x =(1-z_{1})/(1+z_{1})\,, y =(1-z_{2})/(1+z_{2})\,, z =(1-z_{3})/(1+z_{3})$.
In this case, the constraint (\ref{constraintxyz}) takes the following form,
\begin{align}
1+ \sum_{i=1}^{3} \frac{{z_i}^2}{(1-z_{i}^2)}  = 0
\end{align} 
and the symbol alphabet becomes
\begin{align}
\{ z_{i} \,, 1 \pm z_{i}  \,, 1 \pm z_{i} z_{j} \}\,,
\end{align}
for $i,j \in \{1,2,3\}$.
It will be interesting to see whether this more general case is also 
relevant in other situations.

\vspace{0.2 cm}
{\em Acknowledgments.}
It is a pleasure to thank S.~Caron-Huot, H.~Gangl and J.~Zhao for discussions, and T.~Riemann for making the electronic 
form of the results of \cite{Czakon:2004wm} available to us.
J.M.H. was supported in part by the Department of Energy grant DE-SC0009988,
and by the IAS AMIAS fund. 
The work of V.S. was supported by the Russian Foundation for Basic Research through grant
11-02-01196. 

\appendix

\section{Definition of harmonic polylogarithms}
\label{appendix_hpl}

The harmonic polylogarithms  
\cite{Remiddi:1999ew}  
$H_{a_1,a_2,\ldots,a_n}(x)$ 
(HPL), with $a_i \in \{1,0,-1\}$, are defined recursively by
\begin{equation}
H_{a_1,a_2,\ldots,a_n}(x) = \int_0^x  f_{a_1}(t) H_{a_2,\ldots,a_n}(t)\,\dr t\;,
\label{HPL-def}
\end{equation}
where
\bea
f_{\pm 1}(x)&=&\frac{1}{1 \mp x}\;,\;\;f_0(x)=\frac{1}{x}\;,
\label{HPL-f-def}
\\
H_{\pm 1}(x)&=& \mp \log(1\mp x),\;\; H_0 (x)= \log x \;,
\label{HPL-H01-def}
\eea
and at least one of the indices $a_i$ is non-zero.
For all $a_i=0$, one has
\begin{equation}
H_{\underbrace{0,0,\ldots,0}_{n}}(x) = \frac{1}{n!}\log^n x\;.
\end{equation}

\section{Definition of the matrices $B_{i}$}
\label{appendix_bi}

Here we present the non-zero elements of the matrices $B_{i}$ of eq. (\ref{matrixAtilde2loop}).
It is worth noting that all matrices are sparse, i.e. most of their entries are zero.
For brevity of notation, we will denote their entries for each case by $a_{ij}$, hoping that this
will not lead to confusion.
For $B_1$, we have: 
\bea\label{matrixb1}
a_{4,1} &=& -1, a_{4,4} = 1, a_{5,1} = 1, a_{5,5} = 4, a_{5,6} = 3, \
a_{6,5} = -2, a_{6,6} = -1, a_{7,3} = -2, 
\nonumber\\
a_{7,7} &=& 2, a_{8,1} = -2, \
a_{8,3} = -4, a_{8,6} = -4, a_{8,8} = -2, a_{10,4} = 2, a_{10,10} = \
2, a_{11,11} = -6, 
\nonumber\\
a_{12,2} &=& 6, a_{12,12} = 2, a_{12,14} = -8, \
a_{13,13} = -2, a_{14,8} = 3/2, a_{14,12} = 1/2, a_{15,6} = -2, \
\nonumber\\
a_{15,8} &=& 1, a_{15,15} = -1, 
a_{15,16} = 4, a_{16,4} = 1, a_{16,5} = \
-1, a_{16,7} = 1, a_{16,8} = 1, 
\nonumber\\
a_{16,15}&=& -1/2, a_{16,16} = 2, \
a_{19,2} = 3, a_{20,13} = 4, a_{21,1} = -4, a_{21,2} = -8, a_{21,3} = \
-16, 
\nonumber\\
a_{21,4}&=& -4, a_{21,6} = -16, a_{21,8} = 4, a_{21,14} = 16, \
a_{22,12} = 4, a_{22,19} = -8, a_{23,2} = 6, 
\nonumber\\
a_{23,3} &=& 4, a_{23,4}  =
-4, a_{23,5} = 4, a_{23,7} = -4, a_{23,8} = 4, a_{23,10} = -4, \
a_{23,14} = -8,  
\nonumber\\
a_{23,15} &=& 2,
a_{23,16} = 4, a_{23,18} = -4, \
a_{23,19} = 8/3, a_{23,21} = -2, a_{23,22} = 1, a_{23,23} = 3 
\,.
\eea
For $B_2$: 
\bea\label{matrixb2}
 a_{4,4} &=& -2, a_{5,5} = -6, a_{7,7} = -4, a_{8,8} = 2, a_{10,10} = \
-4, a_{12,5} = -8, a_{12,12} = -6, 
\nonumber\\
a_{15,15} &=& 2, a_{16,16} = -4, \
a_{19,7} = -6, a_{19,19} = -2, a_{20,20} = -2, a_{21,15} = 8, \
a_{21,21} = -2, 
\nonumber\\
a_{22,10} &=&-16, a_{22,16} = 16, a_{22,22} = -4, \
a_{22,23} = -4, a_{23,10} = 16, a_{23,23} = -4
\,.
\eea
For $B_3$: 
\bea\label{matrixb3}
 a_{5,5} &=& -2, a_{6,6} = 2, a_{8,8} = 2, a_{11,11} = -4, a_{12,12} = \
-2, a_{13,9} = -3, 
\nonumber\\
a_{13,13} &=& 6, a_{15,8} = -2, a_{21,8} = -4 
\,.
\eea
For $B_4$: 
\bea\label{matrixb4}
  a_{2,2} &=& 2, a_{9,2} = -1, a_{9,3} = 2, a_{9,9} = 1, a_{11,11} = -6, \
a_{12,5} = -4, a_{13,1} = 1,
\nonumber\\
a_{13,2} &=& 3/2, a_{13,3} = 3, a_{13,6} = \
3, a_{13,9} = -3/2, a_{13,13} = 4, a_{13,14} = -4, a_{14,9} = 3/2, \
\nonumber\\
a_{14,13} &=& -2, a_{14,14} = 2, a_{17,1} = -2, a_{17,2} = 2, a_{17,17} \
= -2, a_{17,18} = -4, a_{18,3} = -2, 
\nonumber\\
a_{18,9} &=& -1, a_{18,17} = 1, \
a_{18,18} = 2, a_{19,7} = -3, a_{19,19} = 1, a_{20,7} = 2, a_{20,15} \
= 2, 
\nonumber\\
a_{20,19} &=& -2/3, a_{20,20} = 2, a_{20,21} = -1, a_{21,4} = -8, \
a_{21,15} = 4, a_{21,20} = 4, a_{21,21} = -2, 
\nonumber\\
a_{22,10} &=& -8, \
a_{22,16} = 8, a_{22,23} = -2, a_{23,10} = 8 
\,.
\eea
For $B_5$: 
\bea\label{matrixb5}
 a_{9,9} = 2, a_{13,13} = -2, a_{17,17} = 2, a_{20,20} = -2
\,.
\eea
For $B_6$: 
\bea\label{matrixb6}
a_{2,2} &=& -4, a_{9,9} = -4, a_{11,11} = -4, a_{12,12} = -4, a_{13,13} \
= -4, a_{14,14} = -4, 
\nonumber\\
a_{17,17} &=& 2, a_{18,3} = 4, a_{18,18} = -4, \
a_{19,19} = -4, a_{20,20} = -4, a_{21,4} = 8, 
\nonumber\\
a_{21,8} &=& 4, a_{21,21} \
= 2, a_{22,22} = -4, a_{23,2} = 12, a_{23,3} = 8, a_{23,10} = -8, \
\nonumber\\
a_{23,14}&=& -16, a_{23,16} = 8, a_{23,18} = -8, a_{23,22} = 2, \
a_{23,23} = 2   
\,.
\eea
For $B_7$: 
\bea\label{matrixb7}
a_{11,11} &=& 3, a_{12,5} = 4, a_{12,9} = 6, a_{12,12} = 2, a_{12,13} = \
-4, a_{13,5} = 1, 
\nonumber\\
a_{13,9} &=& 3/2, a_{13,12} = 1/2, a_{13,13} = -1, \
a_{19,7} = 3, a_{19,9} = 3, a_{19,19} = 1, 
\nonumber\\
a_{20,2} &=& -3, a_{20,3} = \
-2, a_{20,14} = 4, a_{20,16} = -4, a_{20,18} = 2, a_{20,20} = 1, 
\nonumber\\
a_{21,4} &=& 4, a_{21,8} = -2, a_{21,15} = -4, a_{21,17} = -2, \
a_{21,21} = 1, a_{22,10} = 8, 
\nonumber\\
a_{22,16}&=& -8, a_{22,20} = -4, \
a_{22,22} = 2, a_{22,23} = 2, a_{23,2} = -6, a_{23,3} = -4, 
\nonumber\\
a_{23,10} 
&=& -4, a_{23,14} = 8, a_{23,16} = -4, a_{23,18} = 4, a_{23,20} = 4, 
\nonumber\\
a_{23,22} &=& -1, a_{23,23} = -1  
\,.
\eea
For $B_8$: 
\bea\label{matrixb8}
a_{11,11} &=& 3, a_{12,5} = 4, a_{12,9} = -6, a_{12,12} = 2, a_{12,13} \
= 4, a_{13,5} = -1, 
\nonumber\\
a_{13,9} &=& 3/2, a_{13,12} = -1/2, a_{13,13} = -1, \
a_{19,7} = 3, a_{19,9} = -3, a_{19,19} = 1, 
\nonumber\\
a_{20,2} &=& 3, a_{20,3} = \
2, a_{20,14} = -4, a_{20,16} = 4, a_{20,18} = -2, a_{20,20} = 1, \
\nonumber\\
a_{21,4} &=& 4, a_{21,8} = -2, a_{21,15} = -4, a_{21,17} = 2, a_{21,21} \
= 1, a_{22,10} = 8, 
\nonumber\\
a_{22,16} &=& -8, a_{22,20} = 4, a_{22,22} = 2, \
a_{22,23} = 2, a_{23,2} = -6, a_{23,3} = -4, 
\nonumber\\
a_{23,10} &=& -4, \
a_{23,14} = 8, a_{23,16} = -4, a_{23,18} = 4, a_{23,20} = -4, \
\nonumber\\
a_{23,22} &=& -1, a_{23,23} = -1 
\,.
\eea
For $B_9$: 
\bea\label{matrixb9}
 a_{11,11} = 5 
\,.
\eea
For $B_{10}$: 
\bea\label{matrixb10}
a_{11,1} = 2, a_{11,2} = 4, a_{11,3} = 4, a_{11,6} = 4, a_{11,14} = \
-8, a_{14,11} = 3/2 
\,.
\eea
For $B_{11}$: 
\bea\label{matrixb11}
a_{11,8} = -2, a_{21,11} = 6
\,.
\eea
Finally, for
$B_{12}$, we have: 
\bea\label{matrixb12}
a_{11,9} = -2, a_{11,13} = 4, a_{13,11} = -3/2 
\,.
\eea

\bibliographystyle{JHEP}

\bibliography{hss_arxiv}

\providecommand{\href}[2]{#2}\begingroup\raggedright\begin{thebibliography}{10}

\bibitem{Bonciani:2003cj}
R.~Bonciani, A.~Ferroglia, P.~Mastrolia, E.~Remiddi, and J.~van~der Bij, {\it
  {Planar box diagram for the (N(F) = 1) two loop QED virtual corrections to
  Bhabha scattering}},  {\em Nucl.Phys.} {\bf B681} (2004) 261--291,
  [\href{http://xxx.lanl.gov/abs/hep-ph/0310333}{{\tt hep-ph/0310333}}].

\bibitem{Kotikov:1990kg}
A.~V. Kotikov, {\it {Differential equations method: New technique for massive
  Feynman diagrams calculation}},  {\em Phys. Lett.} {\bf B254} (1991)
  158--164.

\bibitem{Remiddi:1997ny}
E.~Remiddi, {\it {Differential equations for Feynman graph amplitudes}},  {\em
  Nuovo Cim.} {\bf A110} (1997) 1435--1452,
  [\href{http://xxx.lanl.gov/abs/hep-th/9711188}{{\tt hep-th/9711188}}].

\bibitem{Gehrmann:1999as}
T.~Gehrmann and E.~Remiddi, {\it {Differential equations for two-loop
  four-point functions}},  {\em Nucl. Phys.} {\bf B580} (2000) 485--518,
  [\href{http://xxx.lanl.gov/abs/hep-ph/9912329}{{\tt hep-ph/9912329}}].

\bibitem{Argeri:2007up}
M.~Argeri and P.~Mastrolia, {\it {Feynman Diagrams and Differential
  Equations}},  {\em Int.J.Mod.Phys.} {\bf A22} (2007) 4375--4436,
  [\href{http://xxx.lanl.gov/abs/0707.4037}{{\tt arXiv:0707.4037}}].

\bibitem{Smirnov:2012gma}
V.~A. Smirnov, {\it {Analytic tools for Feynman integrals}},  {\em Springer
  Tracts Mod.Phys.} {\bf 250} (2012) 1--296.

\bibitem{Bonciani:2004gi}
R.~Bonciani, A.~Ferroglia, P.~Mastrolia, E.~Remiddi, and J.~van~der Bij, {\it
  {Two-loop N(F)=1 QED Bhabha scattering differential cross section}},  {\em
  Nucl.Phys.} {\bf B701} (2004) 121--179,
  [\href{http://xxx.lanl.gov/abs/hep-ph/0405275}{{\tt hep-ph/0405275}}].

\bibitem{Smirnov:2001cm}
V.~A. Smirnov, {\it {Analytical result for dimensionally regularized massive
  on-shell planar double box}},  {\em Phys.Lett.} {\bf B524} (2002) 129--136,
  [\href{http://xxx.lanl.gov/abs/hep-ph/0111160}{{\tt hep-ph/0111160}}].

\bibitem{Heinrich:2004iq}
G.~Heinrich and V.~A. Smirnov, {\it {Analytical evaluation of dimensionally
  regularized massive on-shell double boxes}},  {\em Phys.Lett.} {\bf B598}
  (2004) 55--66, [\href{http://xxx.lanl.gov/abs/hep-ph/0406053}{{\tt
  hep-ph/0406053}}].

\bibitem{Czakon:2004wm}
M.~Czakon, J.~Gluza, and T.~Riemann, {\it {Master integrals for massive
  two-loop bhabha scattering in QED}},  {\em Phys.Rev.} {\bf D71} (2005)
  073009, [\href{http://xxx.lanl.gov/abs/hep-ph/0412164}{{\tt
  hep-ph/0412164}}].

\bibitem{Czakon:2005jd}
M.~Czakon, J.~Gluza, and T.~Riemann, {\it {Harmonic polylogarithms for massive
  Bhabha scattering}},  {\em Nucl.Instrum.Meth.} {\bf A559} (2006) 265--268,
  [\href{http://xxx.lanl.gov/abs/hep-ph/0508212}{{\tt hep-ph/0508212}}].

\bibitem{Czakon:2005gi}
M.~Czakon, J.~Gluza, and T.~Riemann, {\it {On the massive two-loop corrections
  to Bhabha scattering}},  {\em Acta Phys.Polon.} {\bf B36} (2005) 3319--3326,
  [\href{http://xxx.lanl.gov/abs/hep-ph/0511187}{{\tt hep-ph/0511187}}].

\bibitem{Czakon:2006pa}
M.~Czakon, J.~Gluza, and T.~Riemann, {\it {The Planar four-point master
  integrals for massive two-loop Bhabha scattering}},  {\em Nucl.Phys.} {\bf
  B751} (2006) 1--17, [\href{http://xxx.lanl.gov/abs/hep-ph/0604101}{{\tt
  hep-ph/0604101}}].

\bibitem{Czakon:2006hb}
M.~Czakon, J.~Gluza, K.~Kajda, and T.~Riemann, {\it {Differential equations and
  massive two-loop Bhabha scattering: The B5l2m3 case}},  {\em
  Nucl.Phys.Proc.Suppl.} {\bf 157} (2006) 16--20,
  [\href{http://xxx.lanl.gov/abs/hep-ph/0602102}{{\tt hep-ph/0602102}}].

\bibitem{Anastasiou:2006hc}
C.~Anastasiou, S.~Beerli, S.~Bucherer, A.~Daleo, and Z.~Kunszt, {\it {Two-loop
  amplitudes and master integrals for the production of a Higgs boson via a
  massive quark and a scalar-quark loop}},  {\em JHEP} {\bf 01} (2007) 082,
  [\href{http://xxx.lanl.gov/abs/hep-ph/0611236}{{\tt hep-ph/0611236}}].

\bibitem{Czakon:2008ii}
M.~Czakon and A.~Mitov, {\it {Inclusive Heavy Flavor Hadroproduction in NLO
  QCD: The Exact Analytic Result}},  {\em Nucl.Phys.} {\bf B824} (2010)
  111--135, [\href{http://xxx.lanl.gov/abs/0811.4119}{{\tt arXiv:0811.4119}}].

\bibitem{vonManteuffel:2013uoa}
A.~von Manteuffel and C.~Studerus, {\it {Massive planar and non-planar double
  box integrals for light Nf contributions to gg $\to$ tt}},
  \href{http://xxx.lanl.gov/abs/1306.3504}{{\tt arXiv:1306.3504}}.

\bibitem{Actis:2006dj}
S.~Actis, M.~Czakon, J.~Gluza, and T.~Riemann, {\it {Planar two-loop master
  integrals for massive Bhabha scattering: N(f)=1 and N(f)=2}},  {\em
  Nucl.Phys.Proc.Suppl.} {\bf 160} (2006) 91--100,
  [\href{http://xxx.lanl.gov/abs/hep-ph/0609051}{{\tt hep-ph/0609051}}].

\bibitem{Actis:2007gi}
S.~Actis, M.~Czakon, J.~Gluza, and T.~Riemann, {\it {Two-loop fermionic
  corrections to massive Bhabha scattering}},  {\em Nucl.Phys.} {\bf B786}
  (2007) 26--51, [\href{http://xxx.lanl.gov/abs/0704.2400}{{\tt
  arXiv:0704.2400}}].

\bibitem{Actis:2007pn}
S.~Actis, M.~Czakon, J.~Gluza, and T.~Riemann, {\it {Fermionic NNLO
  contributions to Bhabha scattering}},  {\em Acta Phys.Polon.} {\bf B38}
  (2007) 3517--3528, [\href{http://xxx.lanl.gov/abs/0710.5111}{{\tt
  arXiv:0710.5111}}].

\bibitem{Actis:2007fs}
S.~Actis, M.~Czakon, J.~Gluza, and T.~Riemann, {\it {Virtual hadronic and
  leptonic contributions to Bhabha scattering}},  {\em Phys.Rev.Lett.} {\bf
  100} (2008) 131602, [\href{http://xxx.lanl.gov/abs/0711.3847}{{\tt
  arXiv:0711.3847}}].

\bibitem{Actis:2007zz}
S.~Actis, T.~Riemann, M.~Czakon, and J.~Gluza, {\it {Two-loop heavy fermion
  corrections to Bhabha scattering}},  {\em eConf} {\bf C0705302} (2007) TEV02.

\bibitem{Actis:2008br}
S.~Actis, M.~Czakon, J.~Gluza, and T.~Riemann, {\it {Virtual Hadronic and
  Heavy-Fermion $O(\alpha^2)$ Corrections to Bhabha Scattering}},  {\em
  Phys.Rev.} {\bf D78} (2008) 085019,
  [\href{http://xxx.lanl.gov/abs/0807.4691}{{\tt arXiv:0807.4691}}].

\bibitem{Henn:2013pwa}
J.~M. Henn, {\it {Multiloop integrals in dimensional regularization made
  simple}},  {\em Phys. Rev. Lett.} {\bf 110} (2013)
  [\href{http://xxx.lanl.gov/abs/1304.1806}{{\tt arXiv:1304.1806}}].

\bibitem{Henn:2013tua}
J.~M. Henn, A.~V. Smirnov, and V.~A. Smirnov, {\it {Analytic results for planar
  three-loop four-point integrals from a Knizhnik-Zamolodchikov equation}},
  \href{http://xxx.lanl.gov/abs/1306.2799}{{\tt arXiv:1306.2799}}.

\bibitem{Chetyrkin:1981qh}
K.~G. Chetyrkin and F.~V. Tkachov, {\it {Integration by Parts: The Algorithm to
  Calculate beta Functions in 4 Loops}},  {\em Nucl.Phys.} {\bf B192} (1981)
  159--204.

\bibitem{Chen1997}
K.-T. Chen, {\it {Iterated path integrals}},  {\em Bull. Amer. Math. Soc.} {\bf
  83, Number 5} (1997) 831--879.

\bibitem{iterated1}
F.~Brown, {\it {Iterated integrals in quantum field theory}},
  \href{http://xxx.lanl.gov/abs/http://www.math.jussieu.fr/~brown/}{{\tt
  http://www.math.jussieu.fr/~brown/}}.

\bibitem{iterated2}
J.~Zhao, {\it {Multiple polylogarithms}},
  \href{http://xxx.lanl.gov/abs/http://www.maths.dur.ac.uk/events/Meetings/LMS/2013/PNTPP13/logzeta.pdf}{{\tt
  http://www.maths.dur.ac.uk/events/Meetings/LMS/2013/PNTPP13/logzeta.pdf}}.

\bibitem{Goncharov:1998kja}
A.~B. Goncharov, {\it {Multiple polylogarithms, cyclotomy and modular
  complexes}},  {\em Math.Res.Lett.} {\bf 5} (1998) 497--516,
  [\href{http://xxx.lanl.gov/abs/1105.2076}{{\tt arXiv:1105.2076}}].

\bibitem{Smirnov:2008iw}
A.~V. Smirnov, {\it {Algorithm FIRE -- Feynman Integral REduction}},  {\em
  JHEP} {\bf 0810} (2008) 107, [\href{http://xxx.lanl.gov/abs/0807.3243}{{\tt
  arXiv:0807.3243}}].

\bibitem{Smirnov:2013dia}
A.~V. Smirnov and V.~A. Smirnov, {\it {FIRE4, LiteRed and accompanying tools to
  solve integration by parts relations}},
  \href{http://xxx.lanl.gov/abs/1302.5885}{{\tt arXiv:1302.5885}}.

\bibitem{Bauer:2000cp}
C.~W. Bauer, A.~Frink, and R.~Kreckel, {\it {Introduction to the GiNaC
  framework for symbolic computation within the C++ programming language}},
  \href{http://xxx.lanl.gov/abs/cs/0004015}{{\tt cs/0004015}}.

\bibitem{Landau:1959fi}
L.~D. Landau, {\it {On analytic properties of vertex parts in quantum field
  theory}},  {\em Nucl. Phys.} {\bf 13} (1959) 181--192.

\bibitem{Gehrmann:2002zr}
T.~Gehrmann and E.~Remiddi, {\it {Analytic continuation of massless two loop
  four point functions}},  {\em Nucl.Phys.} {\bf B640} (2002) 379--411,
  [\href{http://xxx.lanl.gov/abs/hep-ph/0207020}{{\tt hep-ph/0207020}}].

\bibitem{arXiv:math/0606419}
F.~Brown, {\it {Multiple zeta values and periods of moduli spaces
  $\mathfrak{M}_{0,n}$}},  \href{http://xxx.lanl.gov/abs/0606419}{{\tt
  0606419}}.

\bibitem{2009arXiv0908.2238G}
A.~B. {Goncharov}, {\it {A simple construction of Grassmannian
  polylogarithms}},  {\em ArXiv e-prints} (Aug., 2009)
  [\href{http://xxx.lanl.gov/abs/0908.2238}{{\tt arXiv:0908.2238}}].

\bibitem{Remiddi:1999ew}
E.~Remiddi and J.~A.~M. Vermaseren, {\it {Harmonic polylogarithms}},  {\em Int.
  J. Mod. Phys.} {\bf A15} (2000) 725--754,
  [\href{http://xxx.lanl.gov/abs/hep-ph/9905237}{{\tt hep-ph/9905237}}].

\bibitem{Gehrmann:2000zt}
T.~Gehrmann and E.~Remiddi, {\it {Two-Loop Master Integrals for $\gamma^* \to
  3$ Jets: The planar topologies}},  {\em Nucl. Phys.} {\bf B601} (2001)
  248--286, [\href{http://xxx.lanl.gov/abs/hep-ph/0008287}{{\tt
  hep-ph/0008287}}].

\bibitem{Gehrmann:2001ck}
T.~Gehrmann and E.~Remiddi, {\it {Two loop master integrals for $\gamma^* \to
  3$ jets: The Nonplanar topologies}},  {\em Nucl.Phys.} {\bf B601} (2001)
  287--317, [\href{http://xxx.lanl.gov/abs/hep-ph/0101124}{{\tt
  hep-ph/0101124}}].

\bibitem{Binoth:2003ak}
T.~Binoth and G.~Heinrich, {\it {Numerical evaluation of multiloop integrals by
  sector decomposition}},  {\em Nucl.Phys.} {\bf B680} (2004) 375--388,
  [\href{http://xxx.lanl.gov/abs/hep-ph/0305234}{{\tt hep-ph/0305234}}].

\bibitem{Borowka:2012yc}
S.~Borowka, J.~Carter, and G.~Heinrich, {\it {Numerical Evaluation of
  Multi-Loop Integrals for Arbitrary Kinematics with SecDec 2.0}},  {\em
  Comput.Phys.Commun.} {\bf 184} (2013) 396--408,
  [\href{http://xxx.lanl.gov/abs/1204.4152}{{\tt arXiv:1204.4152}}].

\bibitem{Smirnov:2008py}
A.~V. Smirnov and M.~N. Tentyukov, {\it {Feynman Integral Evaluation by a
  Sector decomposiTion Approach (FIESTA)}},  {\em Comput. Phys. Commun.} {\bf
  180} (2009) 735--746, [\href{http://xxx.lanl.gov/abs/0807.4129}{{\tt
  arXiv:0807.4129}}].

\bibitem{Smirnov:2009pb}
A.~V. Smirnov, V.~A. Smirnov, and M.~Tentyukov, {\it {FIESTA 2: parallelizeable
  multiloop numerical calculations}},  {\em Comput. Phys. Commun.} {\bf 182}
  (2011) 790--803, [\href{http://xxx.lanl.gov/abs/0912.0158}{{\tt
  arXiv:0912.0158}}].

\bibitem{Dixon:2011nj}
L.~J. Dixon, J.~M. Drummond, and J.~M. Henn, {\it {Analytic result for the
  two-loop six-point NMHV amplitude in N=4 super Yang-Mills theory}},  {\em
  JHEP} {\bf 1201} (2012) 024, [\href{http://xxx.lanl.gov/abs/1111.1704}{{\tt
  arXiv:1111.1704}}].

\bibitem{wolfram}
S.~Wolfram, {\it {The History and Future of Special Functions}},
  \href{http://xxx.lanl.gov/abs/http://www.stephenwolfram.com/publications/recent/specialfunctions/}{{\tt
  http://www.stephenwolfram.com/publications/recent/specialfunctions/}}.

\bibitem{Kontsevich}
M.~Kontsevich, {\it { Vassiliev's knot invariants}},  {\em Adv. in Soviet
  Math.} {\bf 16(2)} (1993) 137--150.

\bibitem{LewinPolylogarithms}
L.~Lewin, ed., {\em {Structural Properties of Polylogarithms}}.
\newblock American Mathematical Society, 1991.

\end{thebibliography}\endgroup

\end{document}